\documentclass[useAMS,usenatbib]{mn2e}

\def\msunyr{$\mathrm{M}_{\odot}\,\mathrm{yr}^{-1}$}
\def\kms{$\mathrm{km}\,\mathrm{s}^{-1}$}
\def\cm2{cm$^{-2}$}

\usepackage{graphicx}
\usepackage{wasysym}
\usepackage{url}
\usepackage{txfonts}
\usepackage{natbib}
\usepackage{color}
\usepackage{colortbl}
\usepackage{multirow}

\def   \aj {{\rm {AJ}}}
\def   \araa {{\rm {ARA\&A}}}
\def   \apj {{\rm {ApJ}}}

\def   \aap {{\rm {A\&A}}}

\def   \aaps {{\rm {A\&AS}}}

\def   \mnras {{\rm {MNRAS}}}

\def   \apjl{\rm {ApJL}}
\def   \nat{\rm {Nat.}}
\def   \ssr{\rm {Space Sci. Rev.}}

\title[Investigating the inner discs of Herbig Ae/Be stars]{Investigating the inner discs of Herbig Ae/Be stars with CO bandhead and Br\,$\gamma$ emission\thanks{Based on observations made
    with the ESO Very Large Telescope at the Cerro Paranal Observatory
    under programme IDs 079.C-0725, 084.C-0952A, 087.C-0124A and
    279.C-5031A}}

\author[J.~D.~Ilee et al.]
{\parbox{\textwidth}{J.~D.~Ilee$^{1}$\thanks{E-mail: \texttt{john.ilee@st-andrews.ac.uk}},
J.~Fairlamb$^{2}$,
R.~D.~Oudmaijer$^{2}$,
I.~Mendigut\'ia$^{2}$,
M.~E.~van den Ancker$^{3}$,
S.~Kraus$^{4}$ and
H.~E.~Wheelwright$^{5}$
\vspace{0.4cm}}
\\
\parbox{\textwidth}{
$^{1}$SUPA, School of Physics and Astronomy, University of St Andrews, North Haugh, St Andrews, Scotland, KY16 9SS, UK\\
$^{2}$School of Physics and Astronomy, EC Stoner Building, University of Leeds, Leeds, LS2 9JT, UK\\
$^{3}$European Southern Observatory (ESO), Karl-Schwarzschild-Str. 2, 85748 Garching, Germany\\
$^{4}$School of Physics, University of Exeter, Stocker Road, Exeter EX4 4QL, UK\\
$^{5}$Max-Planck-Institut f\"{u}r Radioastronomie, Auf dem H\"{u}gel 69, 53121, Bonn, Germany\\
}}

\begin{document}

\date{Accepted 2014 September 12. Received 2014 September 11; in original form 2014 August 14}
\pagerange{\pageref{firstpage}--\pageref{lastpage}} \pubyear{2014}

\maketitle

\label{firstpage}

\begin{abstract}
Herbig Ae/Be  stars lie in  the mass range  between low and  high mass
young stars, and  therefore offer a unique opportunity  to observe any
changes  in  the  formation  processes  that  may  occur  across  this
boundary.  This paper presents medium resolution VLT/X-Shooter spectra
of six Herbig Ae/Be stars, drawn from a sample of 91 targets, and high
resolution VLT/CRIRES spectra of five Herbig Ae/Be stars, chosen based
on  the presence  of  CO  first overtone  bandhead  emission in  their
spectra.   The X-Shooter  survey reveals  a low  detection rate  of CO
first overtone emission (7 per  cent), consisting of objects mainly of
spectral type B.  A positive correlation is found between the strength
of the  CO $v=$  2--0 and Br\,$\gamma$  emission lines,  despite their
intrinsic linewidths suggesting a separate kinematic origin.  The high
resolution CRIRES spectra are modelled,  and are well fitted under the
assumption  that the  emission originates  from small  scale Keplerian
discs,  interior  to the  dust  sublimation  radius,  but outside  the
co-rotation radius  of the central  stars.  In addition,  our findings
are in very good agreement for the one object where spatially resolved
near-infrared interferometric studies have also been performed.  These
results suggest  that the  Herbig Ae/Be stars  in question are  in the
process of gaining mass via disc accretion, and that modelling of high
spectral resolution spectra  is able to provide a  reliable probe into
the process  of stellar  accretion in young  stars of  intermediate to
high masses.
\end{abstract}

\begin{keywords}
stars: early-type  -- stars: pre-main sequence --  stars: formation --
stars: circumstellar matter -- accretion, accretion discs
\end{keywords}

\section{Introduction}
\label{sec:intro}

Circumstellar discs surrounding young stellar objects (YSOs) have been
the  focus of  much  research because  not only  do  they provide  the
location  for possible  planet formation  to occur,  but they  play an
essential role in  the regulation and evolution of  the accretion that
takes  place during  the star  formation  process (see  the review  of
\citealt{turner_2014}).   Pre-main sequence  Herbig  Ae  and Be  stars
(HAeBes,  see \citealt{waters_1998})  lie  in the  mass range  between
lower mass T Tauri stars ($M_{\star}<2$\,M$_{\odot}$) and short-lived,
obscured      massive      young     stellar      objects      (MYSOs,
$M_{\star}>8$\,M$_{\odot}$).  Thus, they offer a unique opportunity to
observe and characterise any  similarities or differences between low-
and high-mass star formation  processes (see \citealt{larson_2003} and
\citealt{mckee_2007} for reviews).  For example, across the mass range
between T Tauri stars and MYSOs, there is evidence for a change in the
mechanism that  transfers material  from the surrounding  natal cloud,
through the disc,  and on to the central protostar.   The mechanism is
thought  to switch  from T  Tauri-like magnetospheric  accretion -  in
which the  disc is truncated  at radial  distances no larger  than the
co-rotation radius and  accretion proceeds to the  stellar surface via
magnetically           channelled          accretion           funnels
\citep{bertout_1989,bouvier_2007}   -   to    some   other,   as   yet
uncharacterised phenomenon  \citep{vink_2003, vink_2005}.   The reason
for  this  possible  change  in accretion  mechanism  is  because  the
interior envelopes  of HAeBes  are thought to  be mostly  radiative in
nature  \citep{hubrig_2009}.    Therefore,  they  lack   the  interior
convection  currents required  to  power strong  magnetic  fields -  a
requirement  for  such  magnetospheric  accretion  to  occur.   Recent
observations show a low detection rate of magnetic fields ($\sim$7 per
cent)  across  a large  sample  of  HAeBes, supporting  this  scenario
\citep{alecian_2013a}.  It is  possible that the lower  mass Herbig Ae
stars  undergo similar  magnetospheric accretion  to that  of T  Tauri
stars \citep{muzerolle_2004, mottram_2007}, but  the situation for the
higher mass Herbig Be stars is not known.

\smallskip

An alternative  to magnetospheric accretion is  direct accretion (also
called  boundary layer  accretion), where  material from  the  disc is
accreted directly  onto the stellar surface, along  the ecliptic plane
\citep{lynden-bell_1974,  bertout_1988,  blondel_2006}.  For  boundary
layer  accretion to  occur,  a  disc-like geometry  would  have to  be
present on  scales smaller than the  co-rotation radius of  the star -
the  location  at which  any  magnetospheric  accretion funnels  would
likely begin to  operate \citep{shu_1994, muzerolle_2003}.  Therefore,
in order to investigate the accretion mechanisms of these young stars,
information  on the  geometry of  inner regions  of  the circumstellar
disc, close to the central star is required.

\smallskip

While it is believed that dust  is responsible for most of the thermal
emission  from   circumstellar  discs,  it  is  likely   that  gas  is
responsible  for the majority  of their  mass.  HAeBes  possess strong
stellar  radiation  fields  compared  to  that  of  their  lower  mass
counterparts.  Because  of this, regions of  their circumstellar discs
close  to  the   central  star  are  likely  to   be  heated  to  high
temperatures.   If  this  temperature  exceeds  the  dust  sublimation
temperature, then the dust in  the disc is destroyed.  This gives rise
to an  inner disc consisting of only  gas, out to the  location of the
dust  sublimation  radius,  on  scales  of a  few  astronomical  units
\citep[see the review of][]{dullemond_2010}.

\smallskip

Direct observations of these regions  of HAeBes are complicated by the
fact that most  objects lie at relatively large  distances.  Thus, the
small sizes of the regions involved mean that imaging is only possible
using  interferometry.  However,  this is  an  observationally complex
task  limited  to  bright  targets  \citep{tatulli_2008,  kraus_2008b,
  wheelwright_2012}.  Therefore, there is much interest in determining
more indirect  observational techniques that can  probe the conditions
close to the central star.

\smallskip

The  most  abundant  molecule  in  circumstellar  discs  is  molecular
hydrogen (H$_{2}$).   However, the  large energies required  to excite
H$_{2}$, low transition probabilities, and atmospheric absorption
  across  the relevant  wavelength range  mean that  thermal emission
from this molecule is difficult to observe, and it is therefore not an
efficient tracer of these regions.  Coupled rotational and vibrational
emission of the next most abundant molecule, CO, offers an alternative
diagnostic.  CO  bandhead emission (also called  overtone emission) is
excited   in   warm  ($T   =$   2500--5000\,K)   and   dense  (n   $>$
$10^{15}$\,cm$^{-3}$) neutral gas - exactly the conditions expected in
the inner  parts of accretion  discs, making this emission  a valuable
probe  of  these  regions  \citep{glassgold_2004}.   Several  previous
investigations have been successful in fitting the CO bandhead spectra
of young stars under the  assumption that the emission originates from
a  gaseous circumstellar  disc \citep{carr_1989,  blum_2004, bik_2004,
  thi_2005, wheelwright_2010, cowley_2012, ilee_2013}.

\smallskip

The  $n =  7$--4 transition  of atomic  hydrogen (H\,{\sc  i}) in  the
Brackett series  (Br\,$\gamma$) occurs  at $\lambda$  = 2.16\,\micron,
and is also excited at  high temperatures ($T\gtrsim 10^{4}$\,K).  The
origin  of  such hydrogen  recombination  emission  is still  unknown.
Several  theories   have  been  proposed,   including;  magnetospheric
accretion  phenomena   \citep{muzerolle_1998a},  inner   disc  regions
\citep{muzerolle_2004}, stellar winds  \citep{strafella_1998} and disk
winds  \citep{ferreira_1997}. \citet{muzerolle_1998b}  found that  the
Br\,$\gamma$   line    luminosity   in   a   sample    of   low   mass
(0.2--0.8\,M$_{\odot}$) T Tauri stars  was tightly correlated with the
accretion  luminosity   as  measured   from  blue   continuum  excess.
\citet{calvet_2004} extended this investigation to YSOs with masses up
to 4\,M$_{\odot}$,  and find good  agreement with the  previous study,
and the  relationship was used  to examine  the accretion rates  of 36
Herbig   Ae  stars   by  \citet{garcia-lopez_2006}.    More  recently,
\citet{mendigutia_2011}  determined  accretion  luminosities  from  38
Herbig  Ae and  Be stars  by  examining the  UV excess  in the  Balmer
discontinuity, and  found a  correlation with  Br\,$\gamma$ luminosity
similar to \citet{calvet_2004}.

\smallskip

This  paper  utilises  a  collection of  Very  Large  Telescope  (VLT)
X-Shooter  and CRIRES  observations  of several  Herbig Ae/Be  objects
based on the detection of CO first overtone bandhead emission in their
spectra.   The observations,  sample  selection  and determination  of
stellar  parameters  are  described  in  Section  \ref{sec:obs}.   The
measured observable  quantities are presented and  analysed in Section
\ref{sec:obsres}.   Modelling  of the  CO  spectra  is discussed,  and
comments on individual objects  are given in Section \ref{sec:modres}.
Discussion of the results from  both sets of observations is presented
in Section \ref{sec:discussion}, and  finally conclusions are outlined
in Section \ref{sec:conclusions}.

\section{Observations \& sample selection}
\label{sec:obs}

The  observations  for  this  investigation were  obtained  using  two
instruments  on  the  ESO  VLT  at  Cerro  Paranal.   High  resolution
$2.3\,\micron$ spectra  of 5 Herbig  Ae/Be stars, targeted  because of
previous  detection  of  CO  first overtone  bandhead  emission,  were
obtained using  the cryogenic spectrograph  CRIRES \citep{kaufl_2008}.
The CRIRES observations of HD 36917, HD 259431 and HD 58647 were taken
on 26  and 27  October 2010.  Using  a slit width  of 0.2  arcsec, the
observations  achieved spectral  resolution of  approximately 80\,000.
The observations of  PDS 37 were taken on 06  June 2007 and originally
published in \citet{ilee_2013}.   Using a slit width of  0.6 arcsec, a
spectral  resolution  of  approximately  30\,000  was  achieved.   The
archival observations  of HD 101412  were taken  on 5 April  2011, and
were originally  published by \citet{cowley_2012}. Using  a slit width
of 0.2  arcsec, this achieved  a spectral resolution of  over 90\,000.
Telluric line removal for all  CRIRES observations was performed using
standard  stars  at comparable  airmasses,  obtained  during the  same
observing run as the science observations.

\smallskip

In addition to  the targeted CRIRES observations,  a medium resolution
spectroscopic survey was performed using the cross-dispersed wide band
spectrograph  X-Shooter \citep{vernet_2011}.   A total  of 91  objects
were  observed in  service mode  between October  2009 and  March 2010
(\citealt{oudmaijer_2011},  Fairlamb  et  al.,  in  prep).   X-Shooter
provides  simultaneous wavelength  coverage  from 300--2480\,nm  using
three spectrograph arms  - UVB, VIS, and NIR.  The  original sample of
91   Herbig  Ae/Be   stars   were  taken   from   the  catalogues   of
\citet{the_1994} and  \citet{vieira_2003}, and were selected  based on
sky  co-ordinates appropriate  for  the observing  semester.  A  small
number  were discarded  due  to insufficient  brightness or  ambiguous
assignment as  a HAeBe star.   This sample  is larger than  most other
published  studies by  a factor  of 2--5.   In addition  to the  large
sample size, the  use of X-Shooter allows comparison  of many spectral
features  from a  single observation,  which is  important given  that
HAeBes  have been  shown  to be  both  photometrically and  spectrally
variable \citep{oudmaijer_2001}.   This paper  utilises data  from the
NIR arm, and deals mainly with the subset of six objects from the full
sample that showed a detection  of CO first overtone bandhead emission
at 2.3\,\micron.

\smallskip

The observations using X-Shooter achieved  a spectral resolution of $R
\sim 8\,000$ ($\Delta \lambda = 0.28$\,nm at $\lambda = 2.3$\,\micron)
using a  slit width  of 0.4\,arcsec.  A  single pixel  element covered
0.06\,nm,  while  a  resolution   element  covered  4.3  pixels.   The
atmospheric   seeing   conditions   in   the   optical   varied   from
1.1--1.6\,arcsec between observations.  The exposure times ranged from
several minutes  for the brightest sources,  up to 30 minutes  for the
faintest  ones.   Nodding  along  the  slit  was  performed  to  allow
background subtraction.  The  data were reduced with  version 0.9.7 of
the ESO  pipeline software \citep{modigliani_2010}, and  verified with
manually reduced data for a  handful of objects to ensure consistency.
The data were of high quality, with signal-to-noise ratios of 100--140
in most cases across the entire sample of HAeBes.

\smallskip

To correct telluric absorption  features within the X-Shooter spectra,
the  ESO software {\sc  molecfit} was  used (Smette  et al.\  2014, in
prep., Kausch  et al.\  2014, in prep.).   The {\sc  molecfit} program
models   the  atmospheric   absorption  above   the   telescope  using
temperature, pressure and humidity  profiles for the observing site, a
radiative transfer  code, and a database of  molecular parameters.  We
used the code to  accurately model the atmospheric absorption features
in  the telluric  observations themselves.   This then  produced model
telluric spectra tuned to the exact atmospheric conditions measured on
the  night of the  observation, but  free from  the effects  of noise.
These model  spectra were then  used to remove telluric  features from
the science  observations, which resulted in a  better correction than
was possible using the standard stars alone.

\smallskip

\begin{figure}
\centering
\includegraphics[width=\columnwidth,angle=0,trim=0 0 0 0,clip]{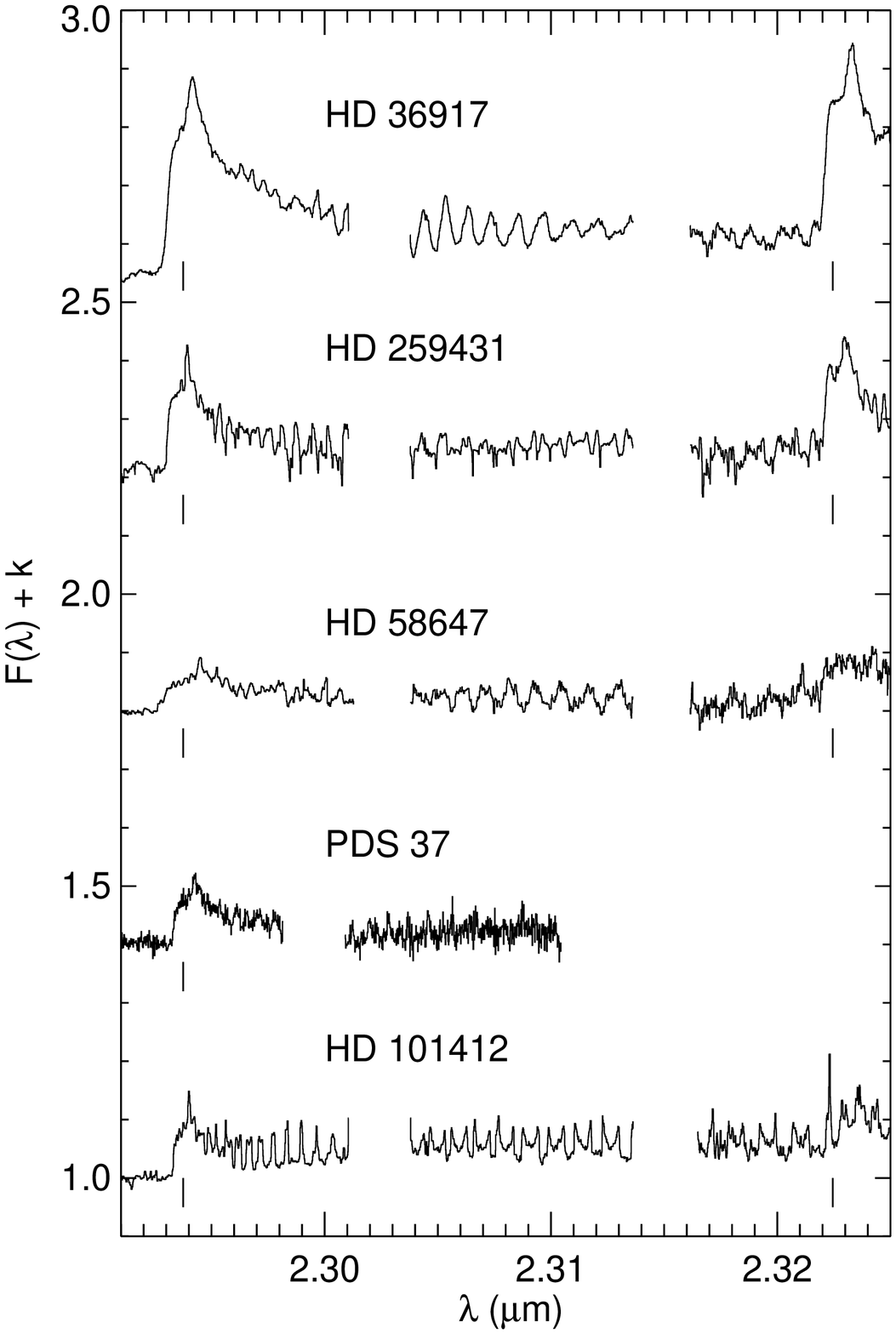}
\caption{VLT/CRIRES spectra of the sample  of objects showing CO first
  overtone  bandhead emission.   Spectra have  been normalised  to the
  continuum and  shifted vertically for clarity.   Vertical ticks mark
  the  rest wavelengths  of the  CO  $v=$ 2--0  and 3--1  transitions,
  respectively. The  observations of  PDS 37  were conducted  using an
  alternative  wavelength setting  to  the rest  of the  observations,
  therefore no data is available beyond 2.311\,$\micron$.}
\label{fig:crires_spectra}
\end{figure}

A  log  of  the  observations  of these  objects  is  shown  in  Table
\ref{tab:haebe_obs}, their  spectra around  the CO first  overtone and
Br\,$\gamma$ region are shown  in Figures \ref{fig:crires_spectra} and
\ref{fig:xshooter_spectra},  and  their astrophysical  parameters  are
given in Table \ref{tab:haebe_stellar_params}.

\smallskip

\begin{table*}
\begin{minipage}{120mm}
\centering
\caption{Log  of  the observations  performed  with VLT/X-Shooter  and
  VLT/CRIRES in which CO first overtone emission was detected.  Signal
  to noise is measured in featureless regions around 2.28\,\micron.}
\label{tab:haebe_obs}
\vspace{1ex}
\small
\begin{tabular}{llllllll}

\hline
Object    & Other name  &RA             & Dec           & Instrument    &  S/N     	&t$_{\mathrm{exp}}$	&  Date      	\\ 
          &             & (J2000)       & (J2000)       &     		&          	& (h)			&               	\\
\hline
HD 36917  & 		& 05:34:47.00	& $-$05:34:10.5 & CRIRES    	&270		& 0.2			& 2010-10-26   	\\
HD 259431 & MWC 147	& 06:33:04.90	& $+$10:19:20.3	& CRIRES    	&172		& 0.2			& 2010-10-26   	\\
HD 58647  & 		& 07:25:56.10	& $-$14:10:45.8	& CRIRES    	&208		& 0.3			& 2010-10-27    \\
PDS 37    & Hen 3$-$373	& 10:10:00.32   & $-$57:02:07.3 & CRIRES    	&114   		& 0.1			& 2007-06-06	\\ 
HD 101412 & V1052 Cen   & 11:39:44.46	& $-$60:10:27.9	& CRIRES	&150		& 0.2			& 2011-04-05    \\
\hline
HD 35929  &		& 05:27:42.79	& $-$08:19:38.6 & X-Shooter	&68		& 0.03			& 2009-12-17	\\		
PDS 133   & SPH 6       & 07:25:04.95  	& $-$25:45:49.7 & X-Shooter	&51	     	& 0.30			& 2010-02-24    \\ 
HD 85567  & V596 Car    & 09:50:28.53   & $-$60:58:03.0 & X-Shooter	&123   		& 0.02			& 2010-03-06    \\ 
PDS 37    & Hen 3$-$373 & 10:10:00.32   & $-$57:02:07.3 & X-Shooter    	&115   		& 0.03			& 2010-03-31    \\ 
HD 101412 & V1052 Cen   & 11:39:44.46   & $-$60:10:27.9	& X-Shooter    	&72    		& 0.06			& 2010-03-30    \\ 
PDS 69    & Hen 3$-$949 & 13:57:44.12   & $-$39:58:44.2 & X-Shooter    	&48    		& 0.03			& 2010-03-29    \\
\hline
\end{tabular}
\end{minipage} 
\end{table*}

\begin{table*}
\begin{minipage}{155mm}
\centering
\caption{Astrophysical parameters of the sample that have been adopted
  for this  work.  Stellar parameters  are determined as  described in
  Section  \ref{sec:stellarparams}  unless   otherwise  stated.   Dust
  sublimation  radii  are  calculated  with  Equation  \ref{eqn:rsub}.
  $K$-band magnitudes are taken from 2MASS \citep{skrutskie_2006}.}
\label{tab:haebe_stellar_params}
\vspace{1ex}
\small
\begin{tabular}{llllllllllll}

\hline
Object    & Spectral    	& K      		&$d$            &$T_{\mathrm{eff}}$	&  $A_{\mathrm{V}}$       & $\log L_{\mathrm{bol}}$  &   $M_{\star}$         & $R_{\star}$   	&	$R_{\mathrm{sub}}$& $v\sin i$	& $R_{\mathrm{cor}}$	\\     
          & Type                &(mags)  		&(pc)           &        (K) 		&         (mags)        &  ($\mathrm{L}_{\odot}$)&($\mathrm{M}_{\odot}$) 	&($\mathrm{R}_{\odot}$) 	&		(au)	& (\kms)	& (au)		    	\\     
\hline
HD 36917  & B9.5e$^{a}$		&  5.7	 		& 470$^{a}$	&  10\,000$^{a}$ 	&  0.5$^{b}$		&   2.20$^{b}$		&     2.5$^{a}$		&   1.8$^{a}$		&		0.4	& 125$^{b}$	& 0.02		     	\\ 
HD 259431 & B6e$^{c}$		&  5.7	 		& 800$^{c}$	&  14\,125$^{c}$	&  1.2$^{c}$		&   3.19$^{c}$		&     6.6$^{c}$		&   6.63$^{c}$		&		3.3	& 100$^{d}$	& 0.07			\\ 
HD 58647  & B9e$^{e}$		&  5.4		 	& 277$^{e}$	&  10\,500$^{f}$	&  1.0			&   2.95$^{f}$		&     3.0$^{e}$		&   2.8$^{e}$		&		0.8	& 118$^{g}$	& 0.03			\\ 
PDS 37    & B2e$^{h}$     	&  7.0  		& 720$^{h}$   	&  22\,000$^{h}$      	&  5.66               	&   3.27                &     7.0       	&   3.0         	&		3.8	& $\dots$	& $\dots$		\\   
HD 101412 & A0III/IVe$^{i}$  	&  7.5  		& $395\pm65$	&  $9\,750\pm250$      	&  0.39               	&   1.58	       	&     2.3         	&   2.2         	&		0.5	& 8$^{l}$	& 0.15			\\       
HD 35929  & F2IIIe$^{o}$	&  6.7			& $325\pm60$	&  $7\,000\pm250$	&  0.10			&   1.67		&     2.7		&   4.6			&		0.5	& 70$^{o}$	& $\dots$		\\
PDS 133   & B6e$^{h}$     	&  9.3  		& $2270\pm500$ 	&  $13\,250\pm1000$	&  1.61               	&   2.16               	&     3.2          	&   2.4         	&		1.0	& $\dots$	& $\dots$ 		\\        
HD 85567  & B7--8Ve$^{m}$ 	&  5.8  		& $470\pm220$ 	&  $12\,500\pm1000$    	&  0.76               	&   2.48                &     3.8          	&   3.9         	&		1.5	& 50$^{q}$	& $\dots$		\\     
PDS 69    & B4Ve$^{n}$    	&  7.2  		& $645\pm120$	&  $16\,500\pm750$      &  1.49               	&   2.81                &     4.7          	&   3.2         	&		2.2	& $\dots$	& $\dots$		\\   
\hline
\end{tabular} 
	\vspace{0.5em}\\ 
      	\footnotesize{$a$: \citet{manoj_2002}}, 
	\footnotesize{$b$: \citet{hamaguchi_2005}},
	\footnotesize{$c$: \citet{kraus_2008a}}, 
	\footnotesize{$d$: \citet{hillenbrand_1992}},
	\footnotesize{$e$: \citet{brittain_2007}},
	\footnotesize{$f$: \citet{montesinos_2009}}.
     	\footnotesize{$g$: \citet{mora_2001}},
	\footnotesize{$h$: \citet{vieira_2003}}, 
	\footnotesize{$i$: \citet{guimaraes_2006}}, 
        \footnotesize{$j$: \citet{dezeeuw_1999}},
        \footnotesize{$k$: \citet{manoj_2006}},
     	\footnotesize{$l$: \citet{vanderplas_2008}},
	\footnotesize{$m$: \citet{vandenancker_1998}}, 
        \footnotesize{$n$: \citet{reipurth_1993}},
        \footnotesize{$o$: \citet{miroshnichenko_2004}},
  	\footnotesize{$p$: \citet{van_leeuwen_2010}},
	\footnotesize{$q$: \citet{miroshnichenko_2001}}.
\end{minipage}
\end{table*}

\subsection{Determining stellar parameters}
\label{sec:stellarparams}

Stellar parameters  for the  targets observed  with CRIRES  were taken
from    various    sources    in    the    literature    (see    Table
\ref{tab:haebe_stellar_params}).  Stellar parameters for the X-Shooter
sample  were  determined  directly  using two  methods  that  will  be
described  in detail  by Fairlamb  et al.\,  (in prep.),  but here  we
briefly summarise the approach.  The first method involved calculating
the  temperature and  surface gravity,  $\log  g$, of  the objects  by
adopting a similar method to  that of \citet{montesinos_2009}.  A best
fit  was performed  across  the hydrogen  recombination  lines of  the
Balmer series  (H\,$\beta$, H\,$\gamma$  and H\,$\delta$)  between the
observed  X-Shooter  spectra  and  a grid  of  Kurucz-Castelli  models
\citep{kurucz_1993,castelli_2004}.  The  fit was made  specifically to
the wings of the recombination  lines, as their broadness is sensitive
to  changes  in  temperature  and  surface  gravity.   Only  the  flux
measurements above  a level  of 0.8 of  the normalised  continuum were
included, to avoid contamination of the fit by any emission component.
These  temperature  and  surface  gravity values  were  then  compared
against the {\sc parsec}  pre-main sequence evolutionary tracks models
\citep{bressan_2012},  which provided  a  corresponding stellar  mass,
radius and luminosity.

\smallskip

However, only half  of the objects in the sample  could be constrained
using this method, due to extreme  emission of the Balmer series which
eclipses even the  broad wings, and often other  strong emission lines
were  present  complicating  any   temperature  estimate.   For  these
objects, known photometry from the  literature was used to determine a
luminosity for the objects.  Then  the Kurucz-Castelli models were fit
to the photometry  by reddening each model until  a best-fitting slope
was found,  providing a temperature  and surface gravity.  A  range of
distances were  then tested  to provide a  luminosity and  radius, and
from the surface gravity a mass was determined.  A cross comparison of
these parameters was then made with the {\sc parsec} tracks, where for
each  temperature and  luminosity pair  there  was a  unique mass  and
radius.  This  then gave  a luminosity  that with  matching parameters
between both the {\sc parsec} tracks  and the photometry fit.  Both of
the methods described  above were tested for consistency  on a handful
of objects, and produced very similar stellar parameters.  The stellar
parameters of  the object PDS  37 proved difficult to  determine using
the  methods above,  so  the  distance and  luminosity  were fixed  to
literature values  \citep{vieira_2003}, which allowed a  mass, radius,
temperature and luminosity to be determined.

\smallskip

Estimation of the typical sizes of important physical regions was also
carried out  for each object -  specifically the location of  the dust
sublimation  radii,  $R_{\mathrm{sub}}$,  and the  co-rotation  radii,
$R_{\mathrm{cor}}$.  One  of the most simple  approaches in estimating
$R_{\mathrm{sub}}$  is   an  analytic  prescription,  such   as  those
described in \citealt{tuthill_2001}  and \citealt{monnier_2002}.  Such
approaches  are   based  on  the  assumptions   about  the  absorption
efficiencies of the  dust in the disc, and using  these assumptions to
calculate  the radius  at which  dust  would survive  given a  certain
stellar luminosity.  However, these  calculations neglect second order
effects, such  back-warming of the disc  material through re-radiation
of  the  stellar heating  from  the  dust  grains,  or the  effect  of
non-homogenously  sized grains.   Addressing such  effects requires  a
proper   treatment   of  radiative   transfer.    \citet{whitney_2004}
calculated a series of  two-dimensional radiative transfer simulations
of  discs  around  young  stars, with  effective  temperatures  up  to
3$\times 10^{4}$\,K.  Their models include the effect of re-radiation,
and use a distribution of dust grain sizes based on the description in
\cite{wood_2002},  with sizes  between 0.01--1000\,\micron.   From the
results  of  these simulations,  the  authors  determined an  analytic
relationship  between  the dust  sublimation  radius  and the  stellar
effective temperature of
\begin{equation}
R_{\mathrm{sub}} = R_{\star} \left( \frac{T_{\mathrm{sub}}}{T_{\star}} \right)^{-2.085},
\label{eqn:rsub}
\end{equation}
where  $T_{\mathrm{sub}}$  is  the   temperature  at  which  the  dust
sublimates.  By  assuming $T_{\mathrm{sub}}$ is 1500\,K,  we have used
this relation  along with the stellar effective  temperatures in order
to  calculate the  dust sublimation  radii of  our objects,  which are
shown  in  Table  \ref{tab:haebe_stellar_params}.  However,  it  seems
worthwhile to  note that it is  difficult to assign a  single value to
$R_{\mathrm{sub}}$  - different  dust species  will likely  survive to
different temperatures based on  their composition, and in reality the
dust sublimation will take place across  a range of radii in the disc.
Nonetheless, as we  are simply estimating the size  of typical regions
within  the  discs,  such  treatment  is  beyond  the  scope  of  this
investigation.

\smallskip

Where  available in  the literature,  measured $v\sin  i$ values  were
obtained and used together with the derived disc inclinations from the
CO  bandhead  fitting  (see  Section \ref{sec:modres}),  in  order  to
determine the stellar  angular velocity $\omega$.  This  was then used
to determine the co-rotation radii,
\begin{equation}
R_{\mathrm{cor}} = \left( \frac{GM_{\star}}{\omega^{2}} \right)^{1/3},
\label{eqn:rcor}
\end{equation}
of     the      objects,     which     are     shown      in     Table
\ref{tab:haebe_stellar_params}.

\section{Observational results}
\label{sec:obsres}

\begin{figure*}
\centering
\includegraphics[width=\textwidth,angle=0,trim=15 0 0 0,clip]{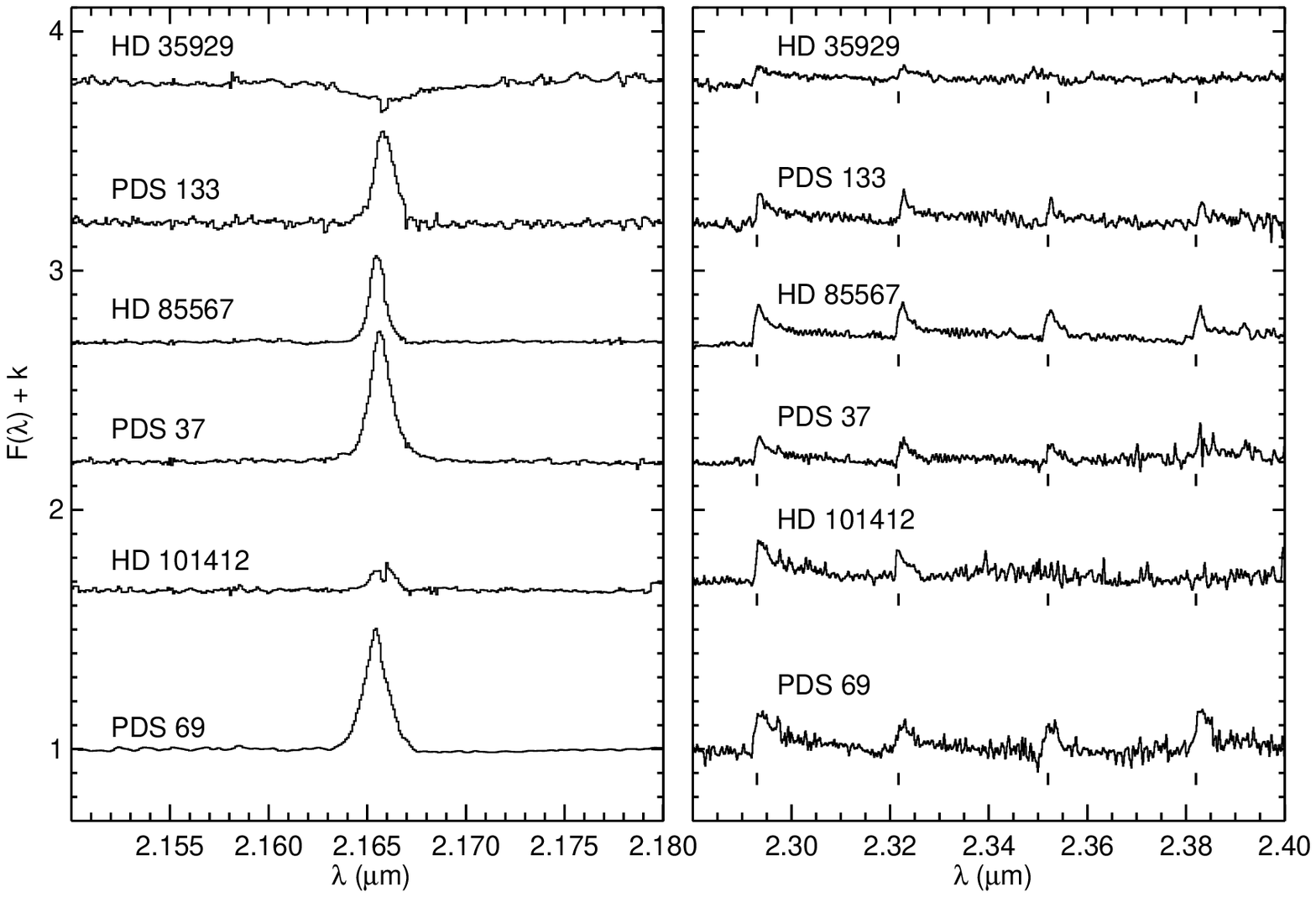}
\caption{VLT/X-Shooter  spectra of  the sample  of objects  from Table
  \ref{tab:haebe_obs}  across the  Br\,$\gamma$  (left)  and CO  first
  overtone (right) wavelength ranges.  Spectra have been normalised to
  their  respective continuum,  shifted  vertically  for clarity,  and
  re-binned by a  factor of 2 to  reduce the effect of  noise.  In the
  right panel, vertical ticks mark the rest wavelengths of the CO $v=$
  2--0, 3--1, 4--2 and 5--3 transitions, respectively.}
\label{fig:xshooter_spectra}
\end{figure*}

From the  sample of  91 Herbig  Ae/Be stars  taken with  X-Shooter, we
investigate two near infrared emission features - Br\,$\gamma$ and the
CO first  overtone bandheads.   The equivalent  widths ($W$)  and full
width at  half maximum (FWHM)  of both  lines were measured  using the
{\sc iraf  noao/onedspec} package.  For each  object, ten measurements
were taken, and an average of  these was reported as the final result,
with the error in this value  given as the standard deviation of these
measurements.  

\smallskip

Because  many A-  and B-type  stars exhibit  photospheric Br\,$\gamma$
absorption, the equivalent widths for the Br\,$\gamma$ lines needed to
be  corrected  for  this  effect.   We adopted  a  method  similar  to
\citet{garcia-lopez_2006}, with the expression
\begin{equation}
W(\mathrm{Br}\,\gamma)_{\mathrm{circ}} = W(\mathrm{Br}\,\gamma)_{\mathrm{obs}} -  W(\mathrm{Br}\,\gamma)_{\mathrm{phot}}\,10^{-0.4\Delta K},
\label{eqn:bry_corr}
\end{equation}
where   $W(\mathrm{Br}\,\gamma)_{\mathrm{obs}}$    is   the   observed
equivalent  width,   $W(\mathrm{Br}\,\gamma)_{\mathrm{phot}}$  is  the
equivalent width of the  corresponding Kurucz-Castelli stellar spectra
based  on spectral  type of  the object,  and $\Delta  K$ is  the disc
continuum emission, computed by subtracting the observed $K$ magnitude
from the photospheric $K$ magnitude  of the corresponding model at the
same given distance.  After performing the correction for photospheric
absorption, Br\,$\gamma$  emission was detected  in 64 objects  in the
sample, absorption  was detected  in 6,  and 21  sources did  not show
detections (where we define non detection for the Br\,$\gamma$ line as
sources which show $-1\,\AA < W_{\mathrm{circ}} < 1\,\AA$).

\smallskip

Six objects exhibited  CO first overtone emission (HD  35929, PDS 133,
HD 85567, PDS 37, HD 101412 and  PDS 69), and the remaining 85 objects
did not  show detections above  a 3-sigma  level across the  $v=$ 2--0
transition.   This  subset  of  six stars  showing  detections  of  CO
overtone  emission in  the X-Shooter  sample  forms the  basis of  the
subsequent  analysis  in  this  work, and  their  spectra  across  the
wavelength range including  Br\,$\gamma$ and CO overtone  are shown in
Figure \ref{fig:xshooter_spectra}.  The objects PDS 133, HD 85567, PDS
37 and  PDS 69 all  show emission from the  $v=$ 2--0, 3--1,  4--2 and
5--3 vibrational transitions, while the objects HD 35929 and HD 101412
only show emission across the $v=$ 2--0 and 3--1 transitions.

\smallskip

The detection rate  of Br$\,\gamma$ emission is 70 per  cent, which is
similar to  other studies of  lower mass T  Tauri stars (74  per cent,
\citealt{folha_2001}).   The  detection  rate  of  CO  first  overtone
emission  is 7  per cent,  which  is lower  than other  studies of  CO
bandhead emission in  young stellar objects, where  detection rates of
around 20 per cent have been reported in low to intermediate mass YSOs
\citep[e.g.][]{carr_1989, connelley_2010} and 17  per cent for massive
YSOs \citep[e.g.][]{cooper_2013}.

\smallskip

The line flux  is calculated from the product of  the equivalent width
of  the  emission  line  (in  the case  of  Br\,$\gamma$  we  use  the
circumstellar  component   $W_{\mathrm{circ}}$)  and   the  extinction
corrected  flux   of  the  object   in  the  K-band  (where   we  take
$A_{\mathrm{K}}  =   0.11\,A_{\mathrm{V}}$,  \citealt{cardelli_1989}).
Examination  of the  near- to  far-infrared continuum  fluxes of  each
object  suggested that  a  extrapolation of  the  K-band magnitude  to
$2.16\,\micron$  and $2.3\,\micron$  is appropriate  to determine  the
line fluxes for both Br\,$\gamma$ and CO.  Line luminosities were then
calculated    using    the    distances    determined    in    Section
\ref{sec:stellarparams},  where  all  measurements  and  corresponding
errors are shown in Table \ref{tab:obs_measured}.

\smallskip

For the CO first overtone emission, the strongest emission relative to
the continuum is detected in PDS 69,  while the weakest is in HD 35929
(the only star of spectral type F  with a CO detection in our sample).
The  strongest  Br\,$\gamma$ emission  relative  to  the continuum  is
observed  in PDS  37, while  the weakest  emission is  detected in  HD
101412.  While most of the  Br\,$\gamma$ emission is single peaked, we
also observe double peaked emission in the spectrum of HD 101412.  The
objects displaying  single peaked  Br\,$\gamma$ emission (PDS  133, HD
85567,  PDS 367  and PDS  69) also  show other  Hydrogen recombination
lines with similar lineshapes - specifically single peaked H\,$\alpha$
and  Pa\,$\gamma$ emission.   HD 101412  shows a  weak double  peak in
H\,$\alpha$  and  Pa\,$\gamma$,  located  within  a  region  of  broad
absorption.  In contrast  to the other objects,  the Br\,$\gamma$ line
in HD 32929  appears to be in  absorption with a very  broad extent in
Figure \ref{fig:xshooter_spectra}, but after correction for the effect
of photospheric absorption, we determine  this line to be in emission.
It  should  be  noted  that  this  object  has  also  previously  been
classified as  a post-main sequence  star \citep{miroshnichenko_2004}.
A full  study of the other  atomic hydrogen lines mentioned  here, and
their diagnostic power, will be  performed in a subsequent publication
(Fairlamb et al.\, in prep.).

\smallskip

Similarly to \citet{brittain_2007}, the lineshapes of the Br\,$\gamma$
emission are broad (130--220\,\kms) and  do not exhibit a blue-shifted
absorption component often seen in  Herbig Ae/Be stars, which suggests
the emission does not originate  in a wind.  The Br\,$\gamma$ emission
of HD 101412 shows a double  peaked line profile, with a separation of
50\,\kms.  As the  source of this double peaking is  unknown, the FWHM
is  measured  across  the  full  width of  the  line,  and  should  be
considered a  maximal value.  The  FWHM of  each lobe of  the emission
corresponds  to approximately  110\,km\,s$^{-1}$.   The  FWHMs of  the
Br\,$\gamma$  emission  are  approximately  5--10  times  the  thermal
linewidth expected for  hydrogen gas at the  effective temperatures of
the  central protostars  (20--30\,\kms).  This  suggests the  hydrogen
recombination lines  are broadened by non-thermal  mechanisms, such as
rotation, turbulence or in-fall toward  the central star.  

\smallskip

\begin{table*}
\begin{minipage}{170mm}
\centering
\caption{Equivalent widths ($W$), line fluxes ($F$), line luminosities
  ($L$) and  line widths  (FWHM) as measured  for Br\,$\gamma$  and CO
  $v=$ 2--0  emission shown  in Figures  \ref{fig:crires_spectra} \&
  \ref{fig:xshooter_spectra}.   Equivalent  widths  for  Br$\gamma$  are
  corrected for photospheric absorption  using the method described in
  Section  \ref{sec:obsres}.  Equivalent  widths for  CO are  measured
  from 2.29--2.30$\,\micron$.  Also shown  is the mass accretion rate,
  $\dot{M}$,  calculated  based  on   the  relationship  described  in
  Equation \ref{eqn:mdot}.}
\label{tab:obs_measured}
\vspace{1ex}
\small
\begin{tabular}{lllllllll}
\hline
Object     	&$W$(CO)  	        &$F$(CO)		        & $L$(CO)            & FWHM(Br\,$\gamma$) 	&$W$(Br\,$\gamma_{\mathrm{circ}}$)	& $F$(Br\,$\gamma$)			& $L$(Br\,$\gamma$)	& $\dot{M}$                             \\
           	&(\AA) 	 		&(W\,m$^{-2}$)			& (L$_{\odot}$)       & (\kms)			& (\AA)					& (W\,m$^{-2}$)	 			& (L$_{\odot}$)		& (\msunyr)				\\
\hline
{\bf CRIRES} \\
HD 36917   	&  $-7.6\pm0.2$		& $1.9\pm0.3\times10^{-15}$	& $-1.9\pm0.3$      & $\dots$			&$\dots$				& $\dots$				& 			&$\dots$				\\
HD 259431  	&  $-4.4\pm0.1$		& $1.2\pm0.2\times10^{-15}$	& $-2.2\pm0.4$       & $\dots$			&$-3.9\pm0.3^{\ast}$			& $1.1\pm0.1\times10^{-15\ast}$		& $-2.1\pm0.4$		&$3.2\times10^{-6}$		\\
HD 58647   	&  $-2.4\pm0.2$		& $7.9\pm0.8\times10^{-15}$	& $-2.2\pm0.1$       & $\dots$			&$-3.3\pm0.3^{\ast}$			& $9.8\pm0.1\times10^{-16\ast}$		& $-1.7\pm0.1$		&$4.8\times10^{-7}$		\\
HD 101412  	&  $-2.8\pm0.3$		& $1.3\pm0.4\times10^{-16}$	& $-3.2\pm0.4$       & $\dots$			&$\dots$				& $\dots$				& 			&$\dots$				\\
PDS 37		&  $-2.0\pm0.4^{\dag}$	& $2.4\pm0.8\times10^{-16}$	& $-2.4\pm0.3$       & $\dots$			&$\dots$				& $\dots$				& 			&$\dots$				\\
{\bf X-Shooter} \\
HD 35929     	&  $-2.4\pm0.4$		& $2.2\pm 0.4\times10^{-16}$	& $-1.7\pm0.2$       & $570\pm90$	        & $-1.3\pm0.9$				& $1.2\pm0.2\times10^{-16}$      	& $-2.6\pm0.2$ 		&$1.5\times10^{-7}$			\\
PDS 133    	&  $-4.0\pm0.7$ 	& $4.0\pm 0.7\times10^{-17}$	& $-2.2\pm0.9$       & $170\pm5$		& $-5.4\pm0.5$				& $5.3\pm0.6\times10^{-17}$		& $-2.2\pm0.9$		&$1.3\times10^{-6}$			\\
HD 85567   	&  $-4.7\pm0.2$		& $1.1\pm 0.1\times10^{-15}$	& $-1.6\pm0.2$       & $130\pm2$		& $-4.3\pm0.1$				& $9.8\pm0.3\times10^{-16}$		& $-1.7\pm0.2$		&$1.2\times10^{-6}$			\\ 
PDS 37     	&  $-3.0\pm0.2$ 	& $3.7\pm 0.3\times10^{-16}$	& $-2.9\pm0.3$       & $190\pm6$	        & $-9.1\pm0.9$				& $1.2\pm0.1\times10^{-15}$		& $-3.2\pm0.5$		&$1.3\times10^{-6}$			\\
HD 101412  	&  $-5.1\pm0.5$ 	& $2.4\pm 0.2\times10^{-16}$	& $-3.1\pm0.4$       & $220\pm20$ 		& $-2.9\pm0.4$				& $1.3\pm0.5\times10^{-16}$		& $-3.4\pm0.4$		&$1.3\times10^{-7}$			\\
PDS 69     	&  $-6.7\pm0.8$		& $4.6\pm 0.6\times10^{-16}$	& $-2.2\pm0.5$       & $205\pm15$		& $-8.4\pm0.6$				& $5.8\pm0.6\times10^{-16}$		& $-2.1\pm0.5$		&$9.1\times10^{-7}$			\\
\hline
\end{tabular} 
\vspace{0.5em}\\ 
	\footnotesize{$\dag$: Data  does not extend across
  	the  full  $v=$  2--0  bandhead,  therefore  this  value  should  be
  	considered a lower limit,}\\
        \footnotesize{$\ast$: taken from \citet{brittain_2007}, where we assume an error of 10 per cent.}
\end{minipage}
\end{table*}

\subsection{A relationship between CO bandhead and Br\,$\gamma$ emission?}

\begin{figure}
\centering   
\includegraphics[width=\columnwidth,angle=0,trim=0 0 0 0,clip]{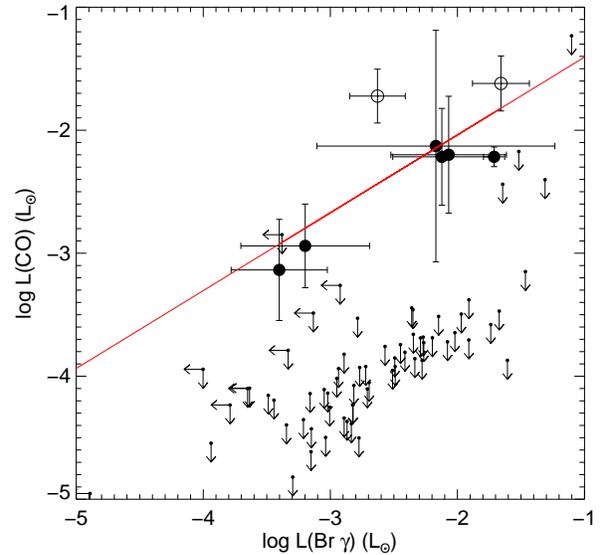}
\caption{A comparison  of the line  luminosities measured from  the CO
  $v=$ 2--0 and Br\,$\gamma$  lines (after correction for photospheric
  absorption) for objects displaying such lines in emission or with an
  upper   limit   for   non-detection.    Filled   circles   represent
  measurements  from  the  simultaneous X-Shooter  observations,  open
  circles  represent  non-simultaneous  measurements from  the  CRIRES
  observations  and literature  data,  and  arrows represent  measured
  upper  limits for  non-detections.  A  best fit  to the  detections,
  $\log L({\mathrm{CO}})  = (0.6\pm0.2)\,\log L({\mathrm{Br}\,\gamma})
  - (0.8\pm0.5)$, is shown with a red line.}
\label{fig:haebe_linelum}
\end{figure}

The relationship between the Br\,$\gamma$  and CO first overtone is of
interest,   as  both   emission  lines   originate  in   circumstellar
environments.  In addition to the simultaneous X-Shooter observations,
the objects  HD 259431 and HD  58647 for which we  present CRIRES data
were    also    shown    to   posses    Br\,$\gamma$    emission    by
\citet{brittain_2007}.    These   fluxes   are   included   in   Table
\ref{tab:obs_measured}.   However, it  should  be  noted that  because
these  data   are  not  simultaneous,  any   effect  of  spectroscopic
variability may alter the values slightly.

\smallskip

Figure \ref{fig:haebe_linelum}  shows the  luminosities of  both lines
measured  from  the X-Shooter  and  CRIRES  datasets.  The  detections
(filled and open circles) occupy the upper ranges of the relationship,
showing that detection of CO  overtone emission is associated with the
detection of  Br\,$\gamma$ emission.   The detections show  a positive
correlation, best fitted with  the relationship $\log L({\mathrm{CO}})
=  (0.6\pm0.2)\,\log L({\mathrm{Br}\,\gamma})  - (0.8\pm0.5)$.   While
this correlation  does not imply  a direct  dependence on both  of the
emission  lines,  it suggests  that  similar  factors may  affect  the
strength of both lines when they are present.

\smallskip

As the calculation  of upper limits for CO first  overtone emission is
complicated by the non-Gaussian shape  of the overall feature, here we
describe the process  adopted to determine them.   Using the modelling
routine described  in Section  \ref{sec:modres}, a synthetic  CO first
overtone spectra  is created, along  with a spectrum of  random noise.
The strength of random noise is altered to produce a series of spectra
that span the  range of SNR shown by the  observations (50--260).  The
strength of  the CO emission is  decreased until the peak  of the $v=$
2--0 bandhead drops below the 3-sigma  level in each of these spectra.
The equivalent width is  then measured across the 2.29--2.3\,$\micron$
wavelength range for  the corresponding model (with  no random noise).
The   relationship  between   the  true   equivalent  width   and  the
signal-to-noise ratio was then determined, and then used to calculated
the upper limit  of equivalent width for each  non-detection, based on
the SNR  of the  spectrum.  The upper  limits for  non-detections were
then  taken  as the  product  of  the  K-band  flux density  and  this
equivalent  width, $F_{\mathrm{UL}}  = F_{\mathrm{K}}W_{\mathrm{UL}}$.
To ensure consistency, we also  perform an identical procedure for the
non detections of Br\,$\gamma$ emission.  However in this case, we use
a  Gaussian lineshape  centered  at 2.1655\,$\micron$,  rather than  a
model bandhead spectra.

\smallskip

Analysis of  the non-detections  shows that the  upper limits  for the
line luminosities of CO first overtone  emission are between 1 and 1.5
dex  below  the luminosities  calculated  from  the detections.   This
suggests  that  our observations  are  not  limited  by noise  in  the
spectra, and that more  sensitive observations with longer integration
times  would not  allow  the  detection of  weaker  CO first  overtone
emission following the same trend.

\smallskip

Given   that  the   line   luminosities  in   both   axis  of   Figure
\ref{fig:haebe_linelum} were calculated  by multiplying the equivalent
widths by the  same continuum fluxes and the same  square distances to
the corresponding objects (Section \ref{sec:stellarparams}), we tested
the  possibility  that  the  correlation  is  spurious.  However,  the
Spearman's  probability of  false correlation  does not  significantly
increase when  both the continuum  and distance values  are considered
through  the partial  correlation  technique \citep{wall_2003}.   This
suggests  that  the  correlation  between  line  luminosities  is  not
spurious, although we note that more  data is necessary to provide any
statistical significance.

\smallskip

This correlation  between the line  luminosities is in  agreement with
several  previous  studies of  both  emission  lines in  young  stars.
\citet{carr_1989} studied  a sample of 40  YSOs, in which they  find a
positive correlation between  the line luminosity of the  CO $v=$ 2--0
emission and  the Br\,$\gamma$  emission in the  10 objects  with such
emission.  \citet{connelley_2010} examined NIR  spectra of 110 Class I
YSOs, and found  a positive correlation between  the equivalent widths
of the  CO $v=$  2--0 and  the Br\,$\gamma$  lines, where  detected in
their sample.

\subsection{Determining accretion rates}
\label{sec:mdot}

As mentioned previously, the luminosity of Br\,$\gamma$ has been shown
to correlate  well with the accretion luminosity  of intermediate mass
YSOs, giving  rise to several correlations between  the two quantities
(see, e.g.\ \citealt{calvet_2004}).  Recently, \citet{mendigutia_2011}
looked  at  a  large  sample  of  Herbig Ae  stars  and  determined  a
correlation following

\begin{equation}
\log \frac{L_{\mathrm{acc}}}{L_{\odot}} = (0.91\pm0.27) \times \log \frac{L_{\mathrm{\mathrm{Br}\,\gamma}}}{L_{\odot}} + (3.55 \pm 0.80).
\label{eqn:mdot}
\end{equation}

We  have  used  this  correlation  between  accretion  luminosity  and
Br\,$\gamma$ line  luminosity to determine  the accretion luminosities
of the  objects possessing Br\,$\gamma$ emission.   Once the accretion
luminosity has been determined, it  is then used to calculate the mass
accretion rate $\dot{M}$ via

\begin{equation}
\dot{M} = \frac{L_{\mathrm{acc}}\, R_{\star}} {G M_{\star}}.
\end{equation}

Table \ref{tab:obs_measured} shows the  mass accretion rates that were
determined  for each object  using this  procedure (including  the two
objects where  measurements were taken  from \citealt{brittain_2007}).
We also  performed measurements  of strength of  Br\,$\gamma$ emission
for the remaining objects in the X-Shooter observations.  

\smallskip

The mass  accretion rates of  the all objects  exhibiting Br\,$\gamma$
emission  span the  range of  10$^{-9}$--10$^{-4}$\,\msunyr,  with the
majority    of    objects    possessing   rates    of    approximately
$1\times10^{-7}$\,\msunyr.   The mass accretion  rates of  the objects
exhibiting  both Br\,$\gamma$ and  CO first  overtone emission  span a
smaller  range of  10$^{-7}$--10$^{-6}$\,\msunyr, with  an  average of
$6\times10^{-7}$\,\msunyr.    Therefore,   while   it  seems   objects
possessing  CO first  overtone emission  span  a small  range in  mass
accretion  rate,  their average  mass  accretion  rate  does not  seem
substantially different to objects without such CO emission.

\smallskip

The   accretion   rates  are   somewhat   higher   than  measured   by
\citet{calvet_2004},  who examined  the spectra  of nine  intermediate
mass  T  Tauri stars  and  find  an  average  mass accretion  rate  of
$3\times10^{-8}$\,\msunyr.   However,  the  HAeBes  studied  here  are
located  at larger  distances, are  more  luminous, and  are thus  (on
average) likely  younger than the  T Tauri stars described  above.  If
mass accretion rates decrease with  stellar age, then this may explain
why the  mass accretion rates of  the HAeBes are higher  than those of
the T  Tauri stars.  When compared  with similar objects, such  as the
study of 38 Herbig Ae  stars by \citet{mendigutia_2011}, the accretion
rates determined here are in agreement with their reported median mass
accretion rate of $2\times10^{-7}$\,\msunyr.

\smallskip

However,  it   should  be   noted  that   the  relation   in  Equation
\ref{eqn:mdot} was  determined from  examination of Herbig  Ae objects
(with $T_{\mathrm{eff}} <  1.2\times10^{4}\,$K), and the applicability
of  this magnetospheric  shock  model  in relation  to  the Herbig  Be
objects    presented     here    has     not    yet     been    proven
\citep{mendigutia_2012}.   Other emission  lines  have  been shown  to
accurately correlate with various  emission excesses, allowing them to
be    used    as    accretion     tracers    (e.g.\    He\,{\sc    i},
\citealt{oudmaijer_2011}).  This  will be  investigated in  detail for
the remainder  of the  X-Shooter dataset  in a  subsequent publication
(Fairlamb et al.\, in prep.).

\section{Modelling the CO bandheads}
\label{sec:modres}

In order  to determine the  origin of the  CO emission, we  fitted the
spectra using  a model describing the circumstellar  environment of CO
as  a  thin  disc   in  Keplerian  rotation,  previously  utilised  in
\citet{wheelwright_2010,  ilee_2013}  and \citet{murakawa_2013}.   The
program is briefly described below.

\smallskip

The population of  the CO rotational levels, to  a maximum of $J=100$,
for  each  $\Delta  v  =  2$  vibrational  transition  considered  are
determined in each cell according to the Boltzmann distribution, which
assumes  local thermodynamic  equilibrium, and  a CO/H$_{2}$  ratio of
10$^{-4}$.  The disc is divided into 75 radial and 75 azimuthal cells.
Each  transition is  assumed  to follow  a  Gaussian with  a width  of
$\Delta \nu$.  The intensity of emission from each cell of the disc is
calculated from  $I_{\nu} = B_{\nu}(T)\left(1-e^{-\tau_{\nu}}\right)$.
The emission is then assigned a weight determined from the solid angle
subtended by  the cell  on the  sky.  The emission  from each  cell is
wavelength shifted  to account for  the line-of-sight velocity  due to
the rotational velocity  of the disc.  The emission  from all cells is
then  summed together,  smoothed to  the instrumental  resolution, and
then shifted in  wavelength to account for the  radial velocity of the
object to produce the entire CO bandhead profile for the disc.

\smallskip

The excitation temperature and surface  number density of the disc are
described analytically as decreasing power laws,
\begin{eqnarray}
T(r) = T_{\mathrm{i}} \left( \frac{r}{R_{\mathrm{i}}} \right)^{p} \\
N(r) = N_{\mathrm{i}} \left( \frac{r}{R_{\mathrm{i}}} \right)^{q},
\end{eqnarray}
where  $T_{\mathrm{i}}$   and  $N_{\mathrm{i}}$  are   the  excitation
temperature  and  surface  density  at  the inner  edge  of  the  disc
$R_{\mathrm{i}}$, and  $p$ and  $q$ are  the exponents  describing the
temperature and  surface density gradient, respectively.   The optical
depth,  $\tau$,  is  taken  to   be  the  product  of  the  absorption
coefficient per CO molecule, and the  CO column density.  Since we are
considering a geometrically thin disc,  the column density is given by
the surface number  density $N$.  The outer radius of  the CO emission
region is taken to be the radius at which $T$ falls below 1000\,K, the
temperature at which  we assume CO overtone emission can  no longer be
excited sufficiently to be detected.

\smallskip

The  best  fitting model  is  determined  using the  downhill  simplex
algorithm,  implemented  by the  \textsc{amoeba}  routine  of the  IDL
distribution.  The  input spectra are first  continuum subtracted, and
then normalised to the peak of the $v=$ 2--0 bandhead.  Model fits are
compared  to  the  data   using  the  reduced  chi-squared  statistic,
$\chi_{r}^{2}$, and the error in the  data is taken to be the standard
deviation  of the  flux in  the pre-bandhead  portion of  the spectra.
Free parameters of  the fit are the inner surface  number density, the
inner  temperature, the  inner  radius, the  intrinsic linewidth,  the
temperature and  density exponents  and the inclination.   The fitting
routine  is repeated  with six  starting positions  spread across  the
parameter   space   to  avoid   recovering   only   local  minima   in
$\chi^{2}_{\mathrm{r}}$,  and   the  final   best  fitting   model  is
determined from these six runs.  The range of parameter space searched
is  shown  in Table  \ref{tab:ranges}.   Errors  on the  best  fitting
parameters are  calculated by holding  all other fitting  variables at
their  best fitting  values, and  altering the  parameter of  interest
until the  difference in  reduced chi-squared,  $\Delta \chi^{2}_{r}$,
increases by unity.

\begin{table}
\centering
\caption{Parameter  space that  is searched  during the  model fitting
  procedure}
\label{tab:ranges}
\begin{tabular}{ll}
\hline
Parameter				&  Range   \\  
\hline
Inclination $i$				&  $0 < i < 90\,\degr$ \\
Intrinsic linewidth $\Delta \nu$	&  $1 < \Delta \nu < 30$\,\kms \\ 
Inner radius $R_{\mathrm{i}}$		&  $1 < R_{\mathrm{i}} < 100$\,R$_{\star}$ \\ 
Inner temperature $T_{\mathrm{i}}$		&  $1000 < T_{\mathrm{i}} < 5000$\,K \\
Inner surface density $N_{\mathrm{i}}$	&  $10^{12} < N_{\mathrm{i}} < 10^{25}$\,cm$^{-2}$ \\ 
Temperature exponent $p$		&  $-4 < p < 0$ \\
Surface density exponent $q$		&  $-4 < q < 0$  \\ 
\hline
\end{tabular}
\end{table}

\subsection{Fitting the X-Shooter observations}

\begin{figure}
\centering
\includegraphics[width=1.0\columnwidth,angle=0,trim=0 0 0 0,clip]{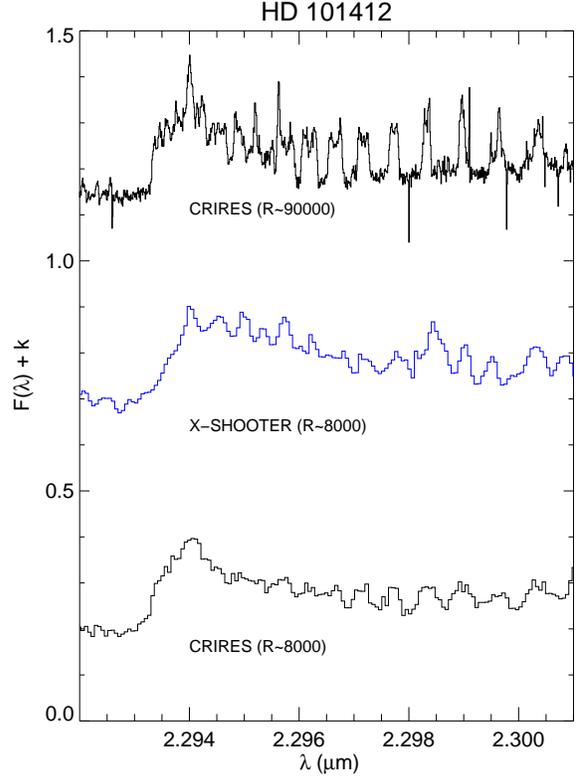}

\caption{Comparison of  CRIRES and X-Shooter data for  HD 101412.  The
  CRIRES spectrum  at $R \sim 90\,000$  shows much detail  that is not
  distinguishable  when  considering  data of  a  lower  spectral
  resolution  of $R  \sim 8\,000$.}
\label{fig:reso}
\end{figure}

The object HD 101412 provides useful test case for our analysis, as it
has been observed with both  X-Shooter at medium resolution and CRIRES
at high resolution.  Figure \ref{fig:reso} shows the comparison of the
$v=$ 2--0 bandhead for both sets of observations.

\smallskip

The CRIRES spectrum at $R \sim  90\,000$ shows much detail that is not
seen when  the considering data of  a lower spectral resolution  of $R
\sim 8\,000$.  In particular, the individual rotational transitions in
the blue  shoulder of  the bandhead  and the  double-peaked rotational
transitions are almost entirely lost  in the X-Shooter spectrum.  When
the CRIRES spectrum is convolved  with a Gaussian corresponding to the
spectral resolution  of the X-Shooter data,  re-binned the appropriate
amount, and  given a similar level  of random noise, the  data appears
qualitatively similar to the X-Shooter spectrum.  However, measurement
of the  equivalent width of  the original and degraded  CRIRES spectra
yielded  very similar  results  (2.8\,\AA), while  measurement of  the
X-Shooter  spectrum  gave  a  larger  equivalent  width  of  5.1\,\AA,
suggesting that  the CO first  overtone emission  in HD 101412  may be
variable.  Nonetheless, it is clear that much information is lost when
the spectrum is degraded to the resolution of X-Shooter.

\smallskip

We performed  the fitting routine  on both the high  resolution CRIRES
data  and the  lower resolution  X-Shooter  data for  HD 101412.   The
fitting  routine performed  poorly on  the X-Shooter  data, recovering
multiple best fit solutions  at similar $\chi^{2}_{\mathrm{r}}$ values
with  very different  parameters  (which  we do  not  show here).   We
attribute  this  to the  low  resolution  data not  showing  important
features in the spectra, such as the blue shoulder of the bandhead and
the narrow,  double peaked  rotational transitions  mentioned earlier.
However, when the fitting routine was performed on the high resolution
CRIRES spectra for HD 101412, a single, unambiguous best fitting model
was obtained.
 
\subsection{Fitting the CRIRES observations}

\begin{figure*}
\centering

	\includegraphics[width=0.97\columnwidth]{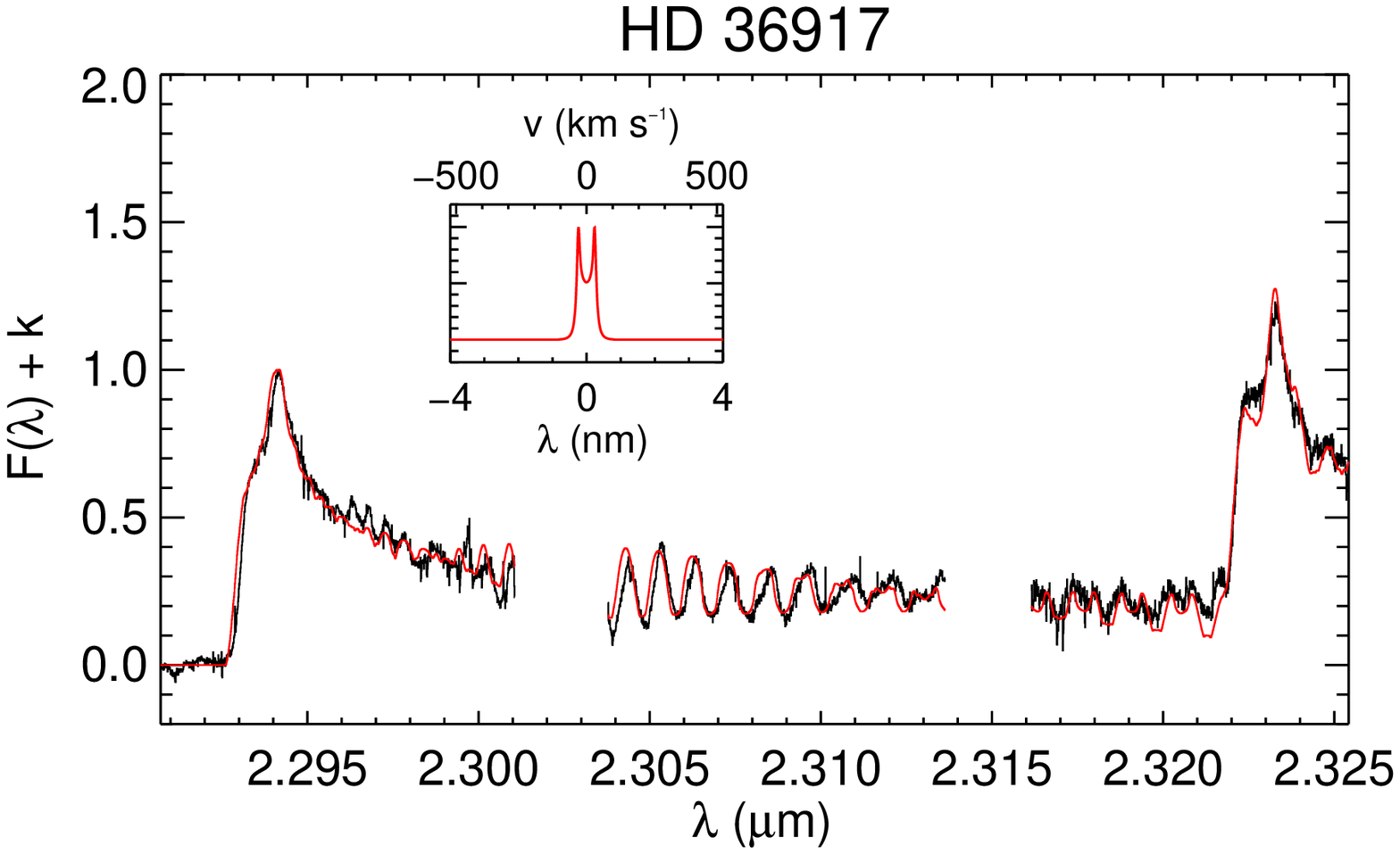}	
	\includegraphics[width=0.97\columnwidth]{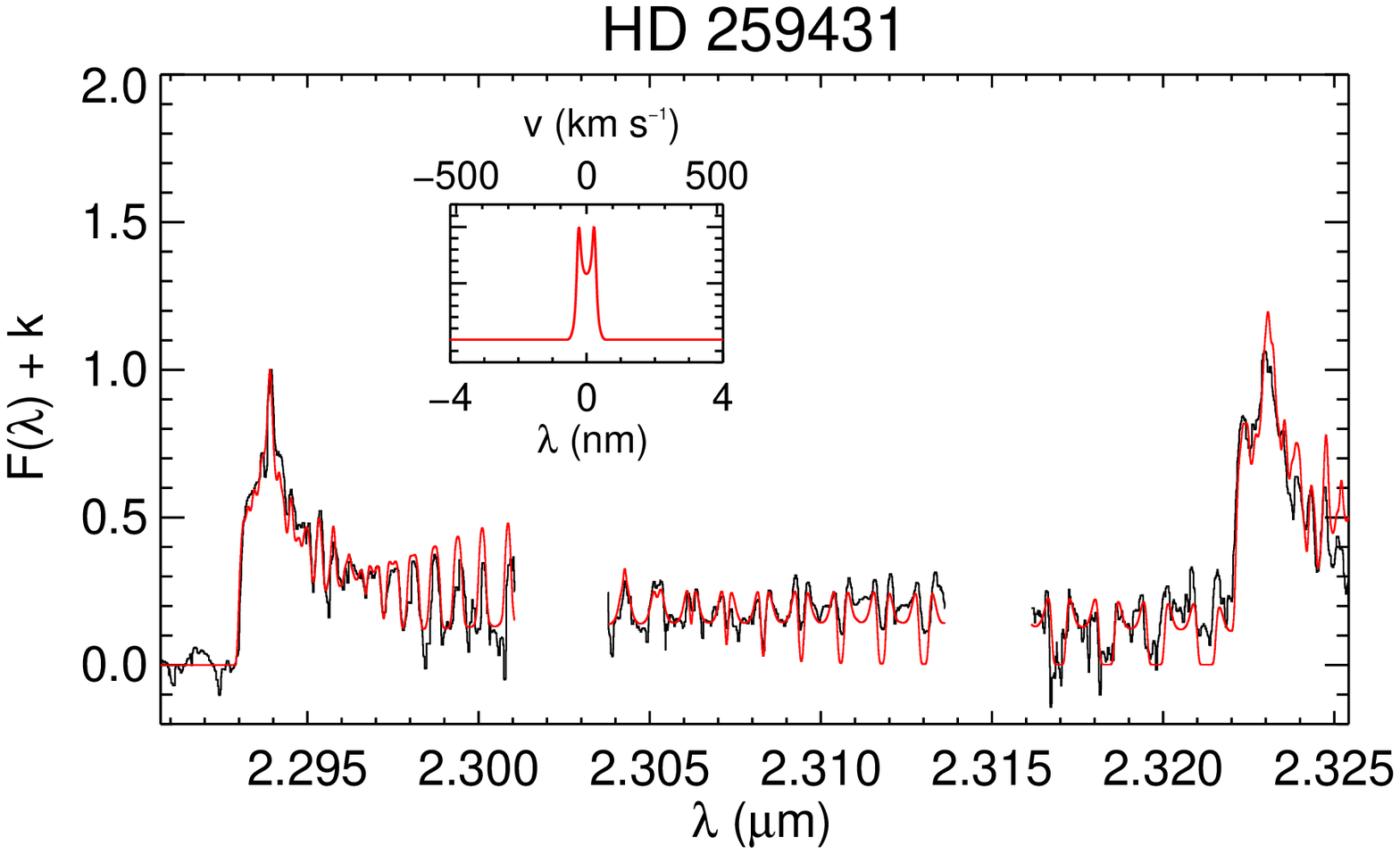}

	\includegraphics[width=0.97\columnwidth]{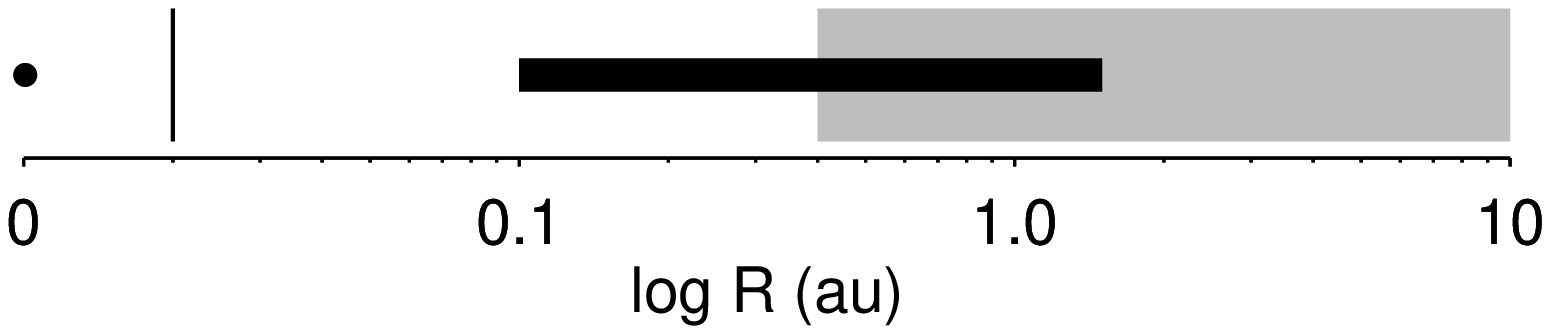}
	\includegraphics[width=0.97\columnwidth]{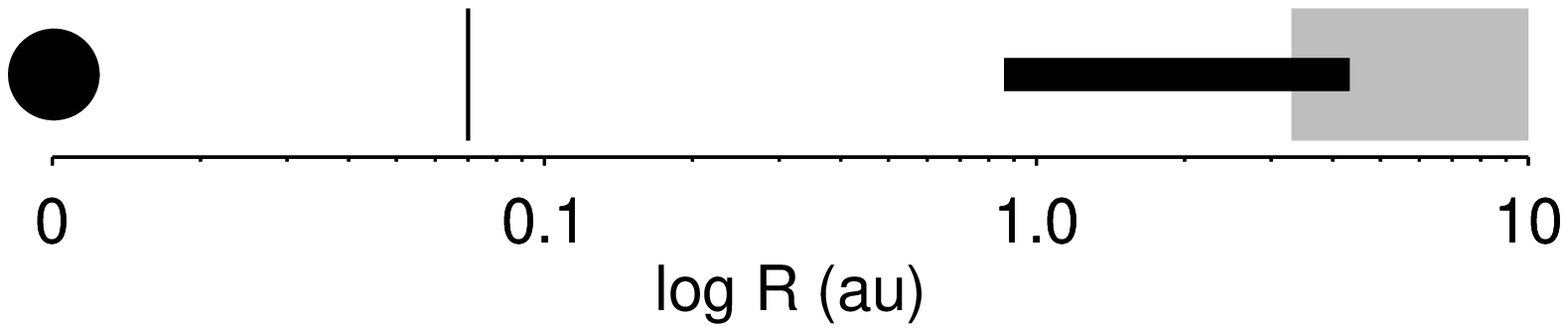}

	\includegraphics[width=0.97\columnwidth]{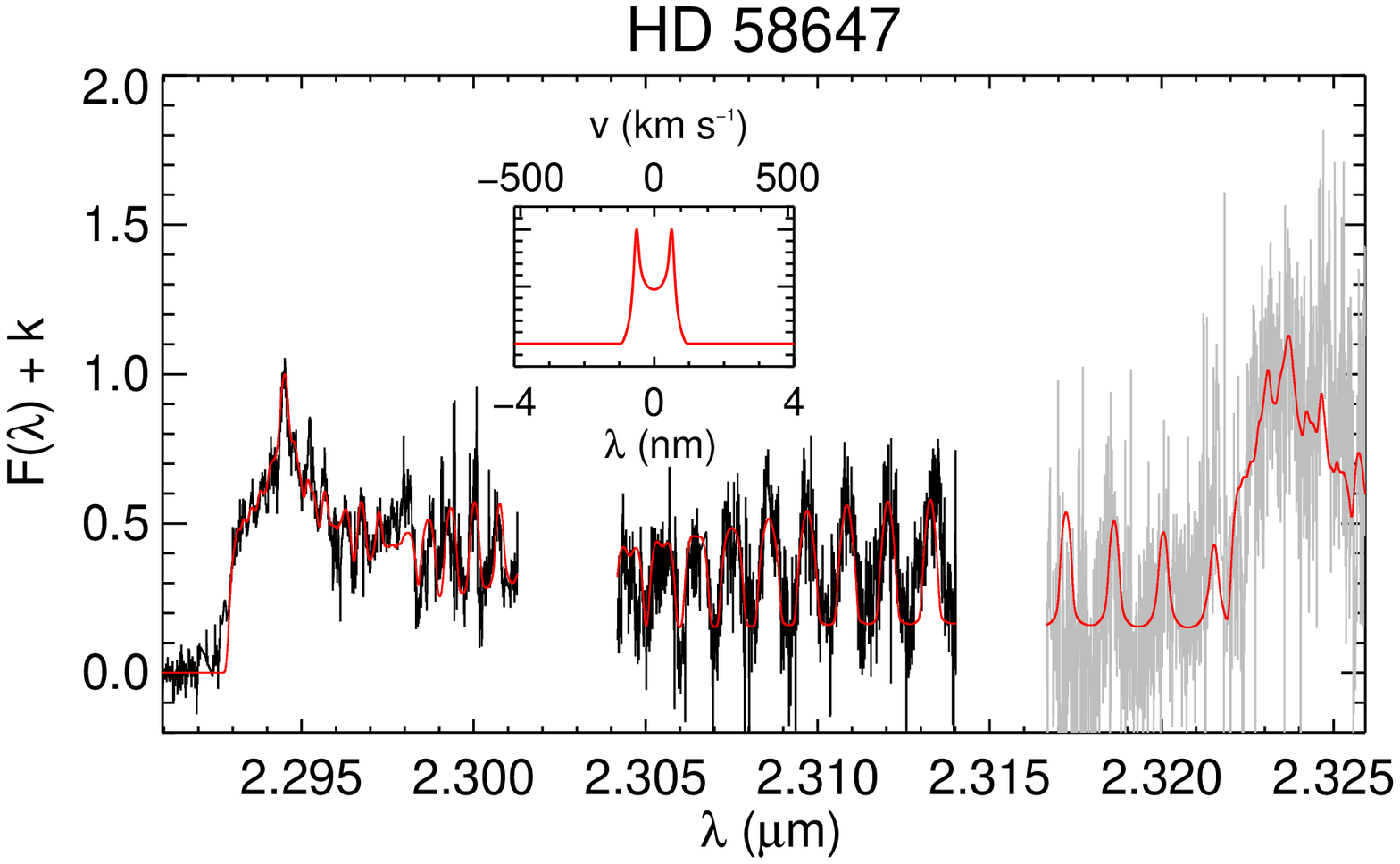}
      	\includegraphics[width=0.97\columnwidth]{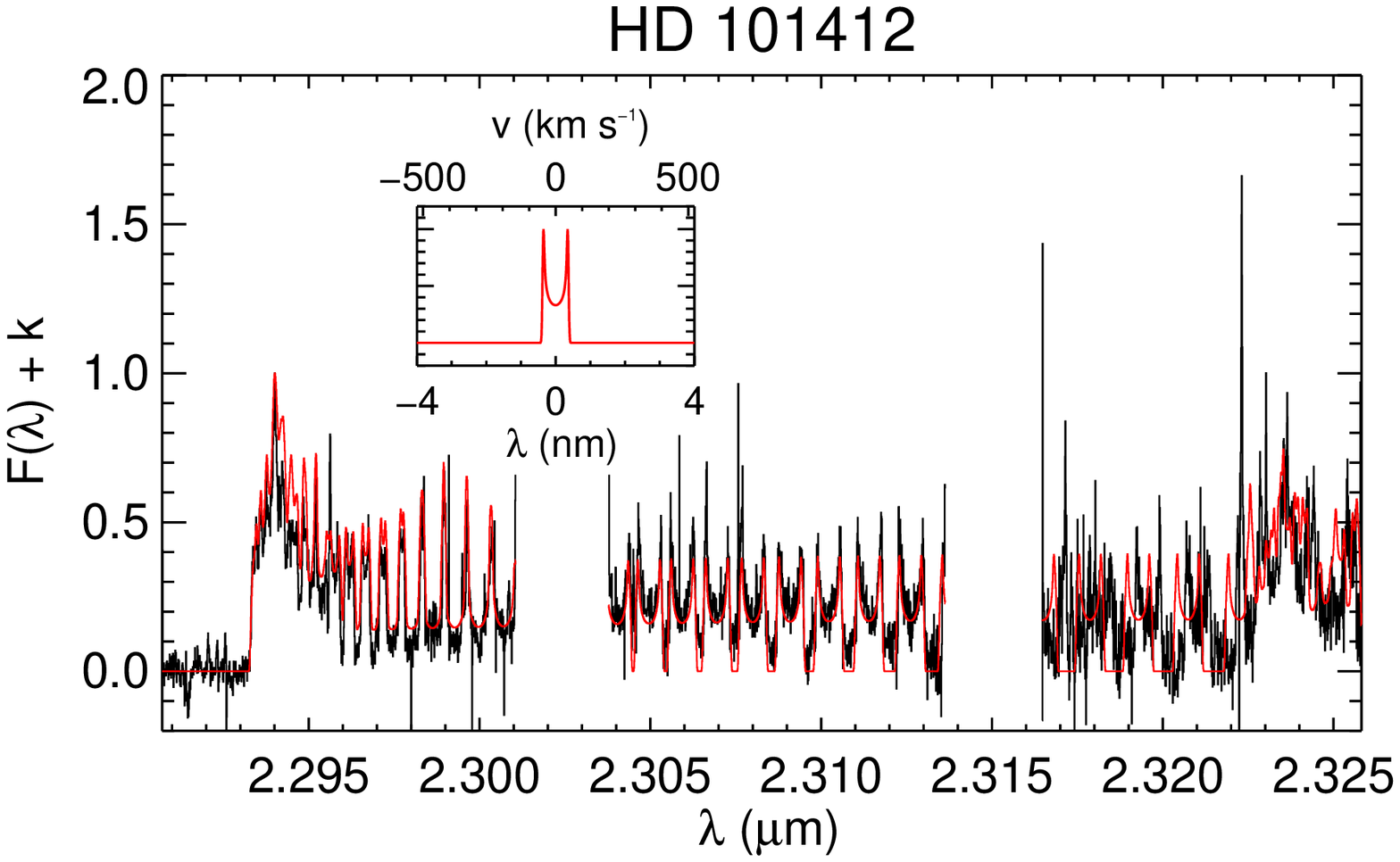}

	\includegraphics[width=0.97\columnwidth]{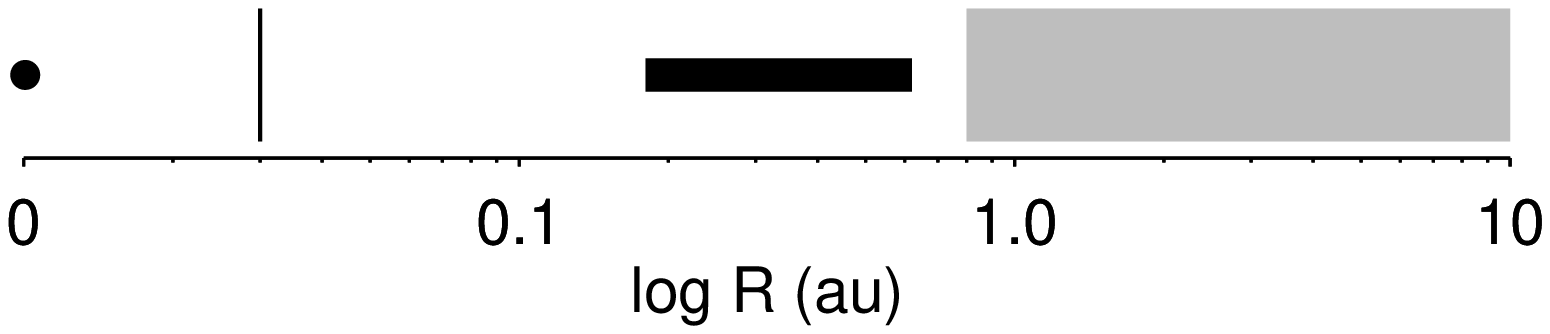}	
 	\includegraphics[width=0.97\columnwidth]{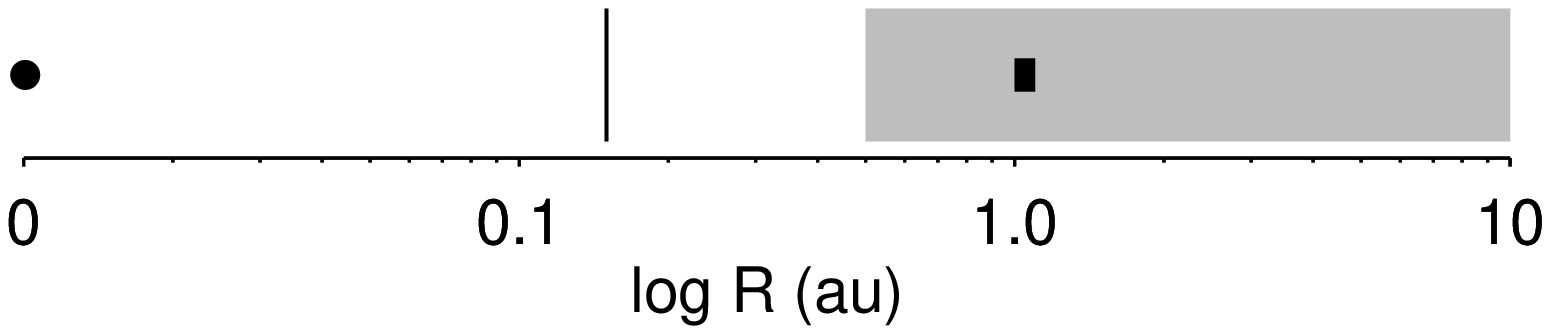}

	\includegraphics[width=0.97\columnwidth]{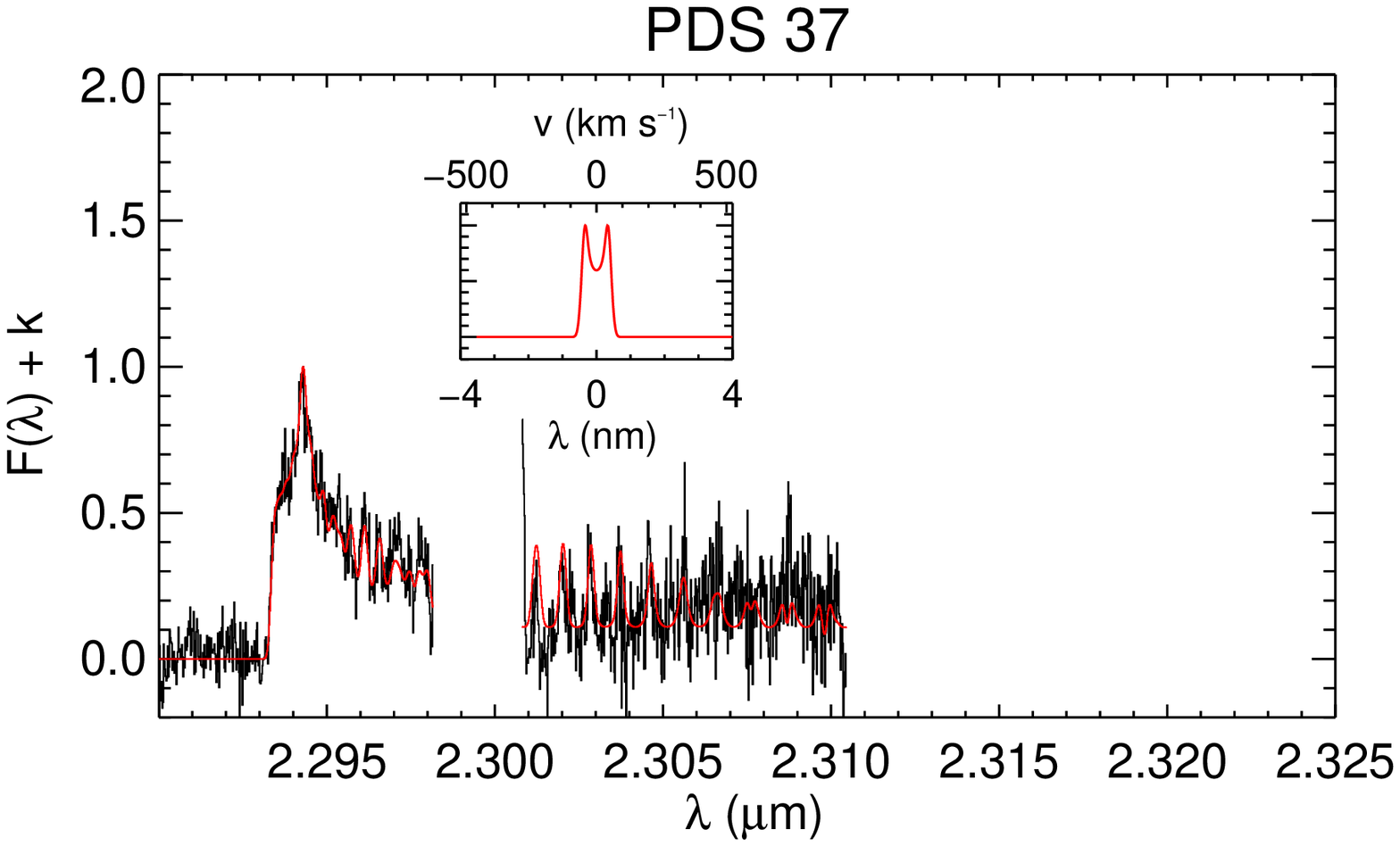}\\
	\includegraphics[width=0.97\columnwidth]{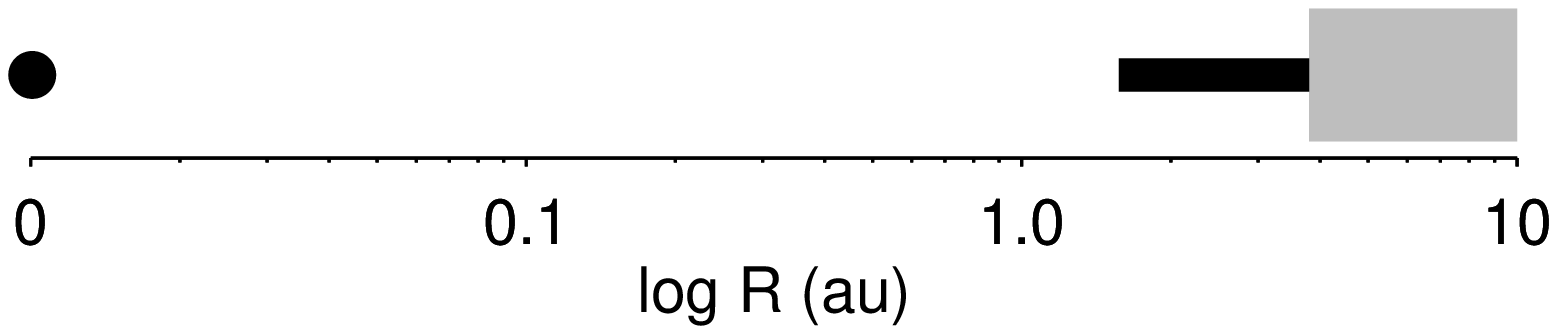}

\caption{CRIRES  spectra  showing  the   CO  first  overtone  bandhead
  emission  (black) with  best fitting  model (red).   Regions of  the
  spectrum excluded from  the fitting are shown in  grey.  Insets show
  the line profile of the $J=$  51--50 transition of CO taken from the
  best fitting model.  Beneath the  spectra are diagrams depicting the
  corresponding structure of  the model: the size of  the central star
  is shown at the origin, the black  region depicts the size of the CO
  emission region, the grey region stretches from the dust sublimation
  radius outwards (Equation \ref{eqn:rsub}), and a vertical line marks
  the  location of  the co-rotation  radius (Equation  \ref{eqn:rcor},
  where data is available).}
   \label{fig:disco_fits}
  \end{figure*}

Given  the   issues  of   fitting  the  X-Shooter   spectra  described
previously, we chose to restrict our  modelling of the CO bandheads to
the 5 objects  observed with CRIRES.  The results of  this fitting are
shown in  Table \ref{tab:disco_fits} and  Figure \ref{fig:disco_fits}.
Below we discuss the fitting results on an object-by-object basis.

\begin{table*}
\begin{minipage}{0.9\textwidth}
\begin{center}
\caption{Best  fitting parameters  obtained from  the fits  to the  CO
  overtone emission using the fitting  routine.  The outer disc radius
  is defined at  the point in the disc in  which the temperature drops
  below 1000\,K, therefore  no error is presented.   Errors shown with
  asterisks  denote  that the  change  in  the  value of  the  reduced
  chi-squared statistic  $\chi^{2}_{r}$ was  less than one  across the
  allowed parameter  range used  in the  fitting procedure  (see Table
  \ref{tab:ranges}).}
\label{tab:disco_fits}
\begin{tabular}{lllllll}
\hline
\noalign{\smallskip} 
Output Parameter &						&HD 36917			&       HD 259431		& 	HD 58647								& HD 101412						& PDS 37	 	\\
\noalign{\smallskip} 
\hline
\noalign{\smallskip} 
CO inner radius 		& $R_{\mathrm{i}}$\,(au)		& 0.1$^{+0.01}_{-0.02}$		&	$0.89^{+0.1}_{-0.1}$	&	$0.18 \pm 0.01$			& $1.0^{+0.2}_{-0.1}$ 				&  1.6$^{+0.1}_{-0.5}$			\\
\noalign{\smallskip} 

CO outer radius 		& $R_{\mathrm{o}}$\,(au)		& 1.5				&	4.3			&	0.62									& 1.1								&  3.8		\\
\noalign{\smallskip} 

Inclination 			& $i$\,(\degr)			&51$^{+7}_{-2}$			&	52$^{+5}_{-3}$		&	$75^{+\ast}_{-10}$   				& $87^{+\ast}_{-20}$					&	89$^{+1\ast}_{-35}$			\\
\noalign{\smallskip} 

Inner surface number density	& $N_{\mathrm{i}}$\,(cm$^{-2}$)	&$6.0^{+2}_{-3}\times10^{20}$ 	&$0.16^{+5.6}_{-\ast}\times10^{20}$			& $0.62^{+12}_{-\ast}\times10^{20}$		& $14^{+75}_{-4}\times10^{21}$   &	0.1$^{+6.3}_{-\ast}\times10^{20}$			\\
\noalign{\smallskip} 

Inner temperature 		& $T_{\mathrm{i}}$\,(K) 		& 3400$^{+800}_{-250}$		&	3200$^{+10}_{-200}$	&	$2800^{+10}_{-300}$	& $1000^{+50}_{-\ast}$				& $5000^{+\ast}_{-1200}$			\\
\noalign{\smallskip} 

Intrinsic linewidth 		& $\Delta \nu$\,(km~s$^{-1}$)	& 5.4$^{+3}_{-2}$ 		& 	10.7$\pm \ast$		&	$10.8 \pm \ast$					& $5.2^{+3}_{-2}$					& $18\pm \ast$				\\
\noalign{\smallskip} 

Temperature exponent 		& $p$				& $-0.5^{+0.1}_{-0.1}$		& 	$-0.74^{+0.4}_{-1.1}$	&	$-0.89^{+0.3}_{-0.9}$	& $ -0.46^{+\ast}_{-2.4}$				& $-1.9^{+0.7}_{-\ast}$		\\	
\noalign{\smallskip} 

Surface number density exponent	& $q$				& $-1.8^{+0.2}_{-0.2}$		& 	$-3.3^{+1.2}_{-\ast}$	&	$-1.6^{+1.5}_{-\ast}$		& $-0.5 \pm \ast$					& $-1.7 \pm \ast$	\\	
\noalign{\smallskip} 
Reduced chi-squared 		& $\chi_{\mathrm{r}}^{2}$		& 3.0				& 	5.3			&	6.2								& 3.8								& 2.5				\\
\noalign{\smallskip} 	
\hline
\end{tabular}  
\end{center}
\end{minipage}
\end{table*}

\subsubsection{HD 36917}

HD 36917  is assumed  to be  a B9.5 type,  2.5\,M$_{\odot}$ star  with a
stellar  radius  of  1.8\,R$_{\odot}$,  an  effective  temperature  of
10$^{4}$\,K and  located at  a distance of  470\,pc \citep{manoj_2002,
  brittain_2007}.  The bolometric luminosity has been determined to be
245\,L$_{\odot}$,   with    a   visual   extinction    of   0.5   mags
\citep{hamaguchi_2005}.

\smallskip

Our modelling of the CO bandheads  indicates a best fitting disc model
extending from 0.1--1.5\,au, at an  inclination of 51\degr.  The inner
edge of the CO emitting region  reaches a temperature of 3400\,K, at a
density of $6\times 10^{20}$\,cm$^{-2}$.   The temperature and surface
number density  exponents are  well constrained  at $-0.5$  and $-1.8$
respectively, and the temperature exponent  agrees well with the value
of  $-0.5$  expected  from  a   flat  disc  in  radiative  equilibrium
\citep{chiang_1997}.  The  location of the CO  emission region crosses
inside  the dust  sublimation radius  of 0.4\,au,  as calculated  from
Equation  \ref{eqn:rsub}, but  lies beyond  the co-rotation  radius of
0.02\,au.  The  intrinsic linewidths of the  individual transitions in
the CO bandhead correspond  to 5.4\,km$^{-1}$, which are approximately
2--5  times the  thermal  linewidths for  CO  at temperatures  between
1000--5000K, indicating broadening by non-thermal mechanisms.

\smallskip

Our results are in contrast to the study of \citet{berthoud_2008}, who
are unable  to fit  their observations  of the  CO $v=$  2--0 bandhead
using a disc and/or ring model, due to the fitting procedure returning
solutions that  converge with unphysical values  (such as temperatures
higher  than  the  dissociation   temperature  of  CO).   The  authors
therefore model  the emission  using an expanding  shell of  CO, which
produces satisfactory fits to the spectrum, in particular the rounded,
convex-shaped  blue  shoulder of  the  bandhead  in their  data.   Our
observations of HD 36917 do not  exhibit such a rounded blue shoulder,
but rather  a traditional concave-shaped blue  shoulder, traditionally
attributed  to emission  from  a disc.   This allows  us  to obtain  a
satisfactory fit to the spectrum using our disc model.  As our data is
of  higher  spectral resolution  than  the  observations presented  in
\citet{berthoud_2008},  it is  unlikely  that the  differences in  the
spectrum  presented  here are  due  to  a  resolution effect.   It  is
possible that the  source of the CO emission in  HD 36917 is variable,
and  that overtones  are excited  both within  a disc  and shell  like
geometry around  the central  star.  However,  time monitoring  of any
spectral variability of the source would be required to confirm this.

\subsubsection{HD 259431}

HD 259431  (MWC 147)  is taken  to be a  6.6\,M$_{\odot}$ star  with a
radius of 6.6\,R$_{\odot}$, having a spectral type of B6, located at a
distance  of  800\,pc.   It   has  a  high  bolometric  luminosity  of
1550\,L$_{\odot}$,   and  an   effective  temperature   of  14\,125\,K
\citep{kraus_2008a}.

\smallskip

The best fitting  disc model for HD 259431  extends from 0.89--4.3\,au
at an  inclination of  52\degr.  It  is interesting  to note  that the
spectro-interferometric  study  of  \citet{kraus_2008a}  determine  an
inclination  of approximately  50\degr\, for  this object,  which agrees
very well our  derived disc inclination.  The  temperature and surface
number  density  of   the  inner  disc  are  3000\,K   and  $2  \times
10^{20}$\,cm$^{-2}$, with  exponents of $-0.7$ and  $-3.3$ respectively.
The temperature  exponent is in  agreement with the expected  value of
$-0.75$  from a  flat blackbody  disc \citep{chiang_1997}.   The inner
edge of  the disc  extends to  within the  dust sublimation  radius of
3.3\,au, however this lies outside  the co-rotation radius of 0.07\,au
as calculated  from $v  \sin i =  100$\,\kms \citep{hillenbrand_1992}.
The  linewidth  of the  individual  rotational  transitions of  CO  is
10.7\,\kms, however this is  not well constrained.  Nevertheless, this
is  a  factor   of  3--8  times  the  thermal  linewidth   for  CO  at
1000--5000\,K.

\smallskip

\citet{brittain_2007} measured the Br\,$\gamma$  emission of HD 259431
and determine a full width at zero intensity (FWZI) of 350\,\kms, with
a luminosity of  $38.1\times10^{-4}$\,$L_{\odot}$ which they calculate
to correspond to a  mass accretion rate of $4.1\times10^{-7}$\,\msunyr
using a relationship based upon  UV veiling.  Using these measurements
with  the   relationship  described  in   \citet{mendigutia_2012},  we
calculate  a higher  mass accretion  rate $3.2\times10^{-6}$\,\msunyr.
\citet{hillenbrand_1992}  also   find  a  higher  accretion   rate  of
$1.01\times10^{-5}$\,\msunyr, determined  from fitting  the SED  of HD
259431.   This discrepancy  may  be  explained by  the  fact that  the
accretion rate calibrations used have not  been proven to be valid for
Herbig Be  stars as  hot as  HD 259431.   

\smallskip

\subsubsection{HD 58647}

The stellar mass of HD 58647  is assumed to be 3.0\,M$_{\odot}$ with a
radius  of   2.8\,M$_{\odot}$,  located  at  a   distance  of  277\,pc
\citep{brittain_2007}.  It has  an effective temperature of 10\,500\,K
and      a     bolometric      luminosity      of     910\,L$_{\odot}$
\citep{montesinos_2009}.

\smallskip

Minor issues were  encountered while fitting HD 58647,  which showed a
decrease in signal-to-noise across  the final detector chip containing
the  $v=$ 3--1  bandhead.  For  this reason,  the fitting  routine was
restricted  to  data from  the  first  and  second detector  chip  and
extrapolated  across the  remaining  data.  Though  this  data in  not
included in the  formal fitting process, it can be  seen that the model
qualitatively  reproduces  the  features  across this  region  of  the
spectrum very well.

\smallskip

The best fitting model  extends from 0.18--0.62\,au, at an inclination
of 75\degr.   This is entirely  within the dust sublimation  radius of
1.1\,au as calculated from Equation \ref{eqn:rsub}.  The inner edge of
the CO emitting region reaches  a temperature of 2800\,K, at a density
of $6.2\times 10^{19}$\,cm$^{-2}$.  The temperature and surface number
density  exponents are  not well  constrained, but  give  best fitting
values of  $-0.89$ and $-1.6$ respectively.   The intrinsic linewidths
of  the  individual  transitions  in  the CO  bandhead  correspond  to
10.8\,\kms, a  factor of  3--8 times the  thermal linewidth for  CO at
1000--5000\,K.

\smallskip

\citet{berthoud_2008}  fitted their  observations of  the CO  overtone
emission in HD 58647 with an  optically thick ring at a temperature of
2380\,K, a  surface density  of $1.6\times10^{20}$\,cm$^{-2}$,  and an
intrinsic linewidth of  7.7\,\kms, seen at an high  inclination to the
line of  sight.  These values  agree with  the fit obtained  using our
observations and disc model to within approximately 1--1.5$\sigma$.
\smallskip

\citet{brittain_2007} measure  the Br\,$\gamma$ emission line  from HD
58647 to be double peaked, with  a full width at zero intensity (FWZI)
of  400\,\kms, and  a  luminosity of  $21.8\times10^{-4}$\,L$_{\odot}$
They  calculate   this  to   correspond  to   an  accretion   rate  of
$3.5\times10^{-7}$\,\msunyr,  which is  consistent with  the accretion
rate   of   $4.8\times10^{-7}$\,\msunyr  determined   using   Equation
\ref{eqn:mdot}.

\subsubsection{HD 101412}
\label{sec:hd101412}

We determine  HD101412 to be  a 2.3\,$\mathrm{M}_{\odot}$ star  with a
radius of  2.2\,$\mathrm{R}_{\odot}$, and an effective  temperature of
9750\,K.    It  has   a  relatively   low  bolometric   luminosity  of
38\,L$_{\odot}$, and  a visual  extinction of $A_{\mathrm{V}}  = 0.39$
mags.  Our modelling of the CO  overtone indicates a best fitting disc
model extending from 1.0--1.1\,au, at  an inclination of 87\degr.  The
inner  edge  of  the  CO  emitting region  reaches  a  relatively  low
temperature of  1000\,K, at  a relatively  high density  of $1.4\times
10^{22}$\,cm$^{-2}$.   The  temperature  and  surface  number  density
exponents are $-0.46$  and $-0.5$ respectively.  These  values are not
well constrained, likely  due to the fact that the  emitting region is
very narrow, and thus determination  of a gradient for the temperature
and density is difficult.  The  intrinsic linewidths of the individual
transitions in the CO  bandhead correspond to 5.2\,km\,s$^{-1}$, which
is 2--4  times the thermal linewidth.   In contrast to all  other best
fitting disc models, the CO emission  region for HD 101412 lies beyond
the  dust  sublimation  radius  of 0.5\,au  calculated  from  Equation
\ref{eqn:rsub}.  Additionally, HD 101412  exhibits the relatively weak
double-peaked  Br\,$\gamma$ emission,  contrary to  the strong  single
peaked Br\,$\gamma$ emission of other objects studied in this work.

\smallskip

HD  101412 was  the subject  of a  similar investigation  involving CO
bandhead emission  by \citet{cowley_2012}.   The authors find  fits to
their spectra  assuming a disc of  CO that is at  most 0.8--1.2\,au in
extent, at a temperature of 2500\,K, and assuming the disc is edge on,
following    the   inclination    of    80\degr\,   determined    from
\citet{fedele_2008}.    Our   results   only  differ   slightly   from
\citet{cowley_2012},  but it  should  be noted  that  their model  and
fitting  routine were different  to the  methods presented  here.  For
instance, they assume  an isothermal ring of CO  with fixed parameters
such as  inclination, and the  fitting was performed visually  with no
systematic  $\chi_{r}^{2}$ minimisation.   However,  the results  here
still suggest a narrow ring  of CO, at approximately distance from the
central protostar, at a slightly cooler temperature.

\smallskip

High  spectral resolution  observations of  the [O\,{\sc  i}] emission
line  at 6300\,{\AA}  were  examined  by \citet{vanderplas_2008},  who
determined that this emission originates from  a region in a disc from
0.15--10\,au, viewed at an inclination  of 30\degr, and corresponds to
a $v\sin\,i$ of 8\,km\,s$^{-1}$.  The authors suggest that HD101412 is
in transition between  a flaring and self shadowed  disc.  The authors
also note a drop in the radial [O\,{\sc i}] emission of 50 per cent at
approximately   0.5\,au  (corresponding   to  their   calculated  dust
sublimation  radius),  and  a  re-brightening  shortly  afterwards  at
approximately 0.8\,au.   The initial  drop is  attributed to  the self
shadowing of a puffed up inner rim at the dust sublimation radius, but
the authors note  that the observed re-brightening  is unexpected.  HD
101412  was  one  of  the  subjects  of  a  study  of  CO  fundamental
ro-vibrational  emission by  \citet{vanderplas_2010}.  The  comparable
linewidths of CO  and [O\,{\sc i}] led this author  to suggest that HD
101412 has a disc with strongly flared gas, but mostly settled dust.

\smallskip

This interpretation could explain why we apparently detect CO bandhead
emission beyond the dust sublimation  radius in HD 101412, in contrast
to the  other objects  studied here.  Our  best fitting model  for the
spectrum of HD 101412 suggests  a relatively cool ($1000$\,K) but high
density ($10^{22}$\,cm$^{-2}$)  emission environment  - if there  is a
sufficient amount of  dense gas located above the  highly settled dust
in the disc, then CO  first overtone emission may originate from these
regions.  The  X-Shooter spectrum  of HD 101412  only exhibits  the CO
$v=$  2--0 and  3--1  bandheads, suggesting  there  is not  sufficient
energy  in the  origin environment  to excite  the  higher vibrational
transitions.  

\smallskip

It is  not clear how our  disc model would perform  when attempting to
fit CO emission that is not  from an axisymmetric disc geometry, which
may explain  why our reported high  inclination is in contrast  to the
low  measured  $v\sin\,i$  of   8\,km\,s$^{-1}$  and  low  inclination
reported  in \citet{vanderplas_2008}.   In  addition, as  there is  an
inversely proportional degeneracy between the location of the emission
and the  disc inclination  in our CO  modelling procedure,  adopting a
lower  inclination would  mean  the corresponding  emission region  is
closer to the star, likely inside the dust sublimation radius.  As the
inclination recovered  for HD 101412  is almost exactly edge  on, then
the reported location of the CO  emission represents an upper limit to
the radial distance of this region.

\subsubsection{PDS 37}

PDS  37 (aka  G282.2988$-$00.7769)  was previously  investigated as  a
massive  young stellar  object in  \citet{ilee_2013}, where  a stellar
mass,  radius  and  effective  temperature  of  11.8\,M$_{\odot}$  and
4.7\,R$_{\odot}$,  26\,100\,K   were  calculated  from   the  bolometric
luminosity and adopted for the  fitting of the CO emission.  This lead
to a best  fitting disc model 1.7--9\,au in  extent, at an inclination
of 80\degr.   The inner temperature of  the disc was  4800\,K, and the
surface density was  $1\times10^{20}$\,cm$^{-2}$, varying with a slope
of  -0.97  and -1.4  respectively.   The  intrinsic  linewidth of  the
transitions was determined to be 16.3\,\kms.

\smallskip

Here we calculate PDS 37 to  have a mass of 7.0\,M$_{\odot}$, a radius
of 3.0\,R$_{\odot}$, a bolometric  luminosity of 1860\,L$_{\odot}$ and
an  effective temperature  of 22\,000\,K.   Modelling the  CO bandhead
emission using these stellar parameters  indicates a best fitting disc
model   extending   from   1.6--3.8\,au,    at   an   inclination   of
$89^{+1}_{-35}$\degr.   The  inner  edge  of the  CO  emitting  region
reaches  a  temperature  of  5000\,K,   at  a  density  of  $1.0\times
10^{19}$\,cm$^{-2}$.   The  temperature  and  surface  number  density
exponents are not well constrained  at $-1.9$ and $-1.7$ respectively.
The location of the CO emission region coincides with dust sublimation
radius of  1.5\,au, as  calculated from Equation  \ref{eqn:rsub}.  The
intrinsic linewidths of the individual  transitions in the CO bandhead
correspond to 18\,\kms, which is approximately 6--14 times the thermal
linewidths for CO at temperatures between 1000--5000\,K.  Altering the
stellar parameters changes the best fitting model parameters slightly,
however still  indicates a relatively  large CO emission region,  at a
high temperature, viewed almost edge-on.

\smallskip

PDS 37 is  also the subject of a spectropolarimetric  study by Ababakr
et  al.\ (in  prep.), where  strong polarisation  signatures are  seen
across the H\,$\alpha$ and doubly  peaked Fe\,{\sc ii} emission lines,
indicating the presence of a gaseous disc viewed at a high inclination
to the line of sight.

\section{Discussion}
\label{sec:discussion}

\subsection{The detection rate of CO first overtone emission}

From an initial sample of  90 targets obtained with X-Shooter (the most
complete spectroscopic sample of Herbig  Ae/Be stars to date), we find
a  low  detection rate  of  CO  first  overtone bandhead  emission  of
approximately 7 per cent.  While a  low detection rate in itself is in
agreement with  previous studies, our detection  rate is substantially
lower than studies  of low mass T Tauri and  Herbig Ae stars \citep[20
  per  cent,][]{carr_1989, connelley_2010},  and also  of  higher mass
MYSOs  \citep[17  per cent,][]{cooper_2013}.

\smallskip

It is also  striking that although our full  X-Shooter sample contains
many A-type stars,  we have only detected CO overtone  emission in one
A-type star. In contrast to this,  although our sample contains few B-
and F-type stars, we have detected CO in a total of 7 B-type stars and
one F-type  star.  So, also in  our sample there may  be evidence that
the  detection  rate for  CO  first  overtone  emission is  lower  for
intermediate-mass young stars than for low- and high-mass young stars.
Below we discuss possible explanations for this.

\smallskip

High temperatures are required to  excite the CO sufficiently in order
for CO bandhead emission to  become detectable. Several of the objects
in our study  are B-type stars (and  HD 101412 is of  type HA0, having
also been  classified as B-type by  \citealt{manoj_2006}).  Therefore,
these  objects  are   hotter  and  more  massive   than  their  A-type
counterparts.  It could  be that many T Tauri and  Herbig Ae stars are
not  hot  enough   to  continually  excite  the  CO   overtones  in  a
circumstellar disc  environment.  In such cases,  variable CO emission
in lower  mass YSOs could be  explained by bursts of  active accretion
\citep{biscaya_1997} or  by originating  from a  different environment
(e.g.\  magnetic   funnel  flows,   \citealt{martin_1997}).   However,
modelling of high  spectral resolution observations of  a large number
of  T Tauri  stars would  be  required to  confirm the  origin of  the
emission.

\smallskip

It is also possible that the majority HAeBes do not have enough gas in
their close circumstellar environments  to allow sufficient excitation
of the CO bandheads.  In addition to high temperatures, high densities
(n $>$  $10^{15}$\,cm$^{-3}$) are required before  this ro-vibrational
emission becomes sufficiently excited  to be detectable.  While direct
measurements of the  amount of gas within these  young stellar systems
is  difficult, there  is evidence  of cleared  gaps around  many HAeBe
stars (a recent example being HD 142527, \citealt{casassus_2012}).  If
the gas  within these  inner regions is  cleared efficiently,  then it
will not be possible to reach  densities high enough to allow overtone
emission to occur.  Our modelling of the CO overtone spectra indicates
the emission originates from environments with a surface density of at
least   $10^{20}$\,cm$^{-2}$.   In   addition,  \citet{muzerolle_2004}
present models of  the inner regions of discs around  HAeBe stars, and
show  that for  accretion rates  greater than  $10^{-8}$\,\msunyr, the
inner  gaseous  disc becomes  optically  thick.   The accretion  rates
determined from our analysis of  the Br\,$\gamma$ in these objects are
above  this level,  further suggesting  that these  objects possess  a
large amount of gas of a high density close to the central protostar.

\smallskip

One  unresolved issue  with this  interpretation is  that there  are a
handful of objects possessing high accretion rates that do not exhibit
CO first  overtone emission.   The work of  \citet{calvet_1991} showed
that CO in absorption could be expected from high mass accretion rates
(observed in FU Ori objects), however we do not detect such absorption
in our observations.

\subsection{The location and orientation of the emitting regions}
\label{sec:location}

The location of the detected emission is of interest, as it determines
which regions of the circumstellar environment can be probed.  We find
that four out of five of  the objects possess best fitting disc models
with  inner   radii  located   interior  to  the   corresponding  dust
sublimation radii.  This suggests that the CO emission originates from
a gaseous  disc, close to the  central protostar.  The one  object for
which this is not the case, HD 101412, exhibits features which are not
typical when compared  to the other objects studied  here (see Section
\ref{sec:hd101412}).

\smallskip

The co-rotation radii lie between  0.02--0.23\,au, and are interior to
the dust sublimation  radius in all objects.  The CO  emission is also
shown  to originate  from beyond  the co-rotation  radius, in  objects
where  $v\sin  i$  measurements   are  available  in  the  literature.
Therefore,  our  modelling  suggests  that  while  CO  first  overtone
emission  is a  valuable probe  of  the inner  gaseous disc  component
around  young stars,  other  spectral tracers  are  required to  trace
regions close  to the  co-rotation radius,  where any  deviations from
magnetospheric accretion geometry are likely to occur.

\smallskip

The  inclinations  of   the  best  fitting  disc   models  range  from
$51$--$72\degr$,  suggesting   a  preference  for  moderate   to  high
inclinations.  While the  number of objects modelled in  this paper is
too low to accurately determine  the statistical significance of this,
it is  nonetheless possible  that a geometric  selection effect  is at
work.  One possible explanation for  a preference toward more inclined
discs could be that  in addition to a inner disc,  the CO emission may
also trace  the vertical inner wall  of the dust disc,  located at the
dust sublimation radius.  However,  further investigation using models
that are able  to include such emission geometry would  be required to
confirm this.

\smallskip

A preference for  moderate to high inclinations is in  contrast to the
study of CO emission of massive YSOs by \citet{ilee_2013}, which found
an essentially random orientation of disc inclinations.  The masses of
the  objects studied  in  \citet{ilee_2013} were  determined from  the
bolometric  luminosity  of  the   objects,  which  may  have  included
contributions from accretion, and  could therefore be overestimates of
the  true stellar  masses.   In  such cases,  an  overestimate of  the
stellar mass  can lead to  a lower inclination being  recovered.  This
effect can  be seen  in our  modelling of PDS  37 -  in this  work, we
recover  a  an  inclination  of   87\degr\,  using  a  stellar  mass  of
7.0\,M$_{\odot}$, while in \citet{ilee_2013} we recover an inclination
of 80\degr\, from a stellar  mass of 12\,M$_{\odot}$.  While this effect
is small, it  may explain why no such preference  for moderate to high
inclination angles was found for MYSOs.

\smallskip

A  positive  correlation  between  the line  luminosities  of  the  CO
bandhead and  Br\,$\gamma$ is found, and  while this does not  imply a
direct dependence (and the number of  objects with emission is too low
to attribute a  statistical significance to the  correlation), it does
suggest  that similar  factors affect  the strength  of both  emission
lines.    However,  analysis   of  the   linewidths  shows   that  the
Br\,$\gamma$  emission  is  approximately  20 times  larger  than  the
corresponding  linewidths   obtained  from  the  fitting   of  the  CO
bandheads.  This difference  in linewidth suggests that  both lines do
not originate  in the  same kinematic  environment, and  are therefore
likely  not   co-spatial.   This   is  in   contrast  to   the  recent
interferometric   study    of   \citet{eisner_2014},   who    find   a
near-coincidence of  CO overtone, Br\,$\gamma$ and  continuum emission
in 5  YSOs.  The spectro-interferometric study  of \citet{kraus_2008b}
suggests  two possible  origins  for Br\,$\gamma$  emission -  compact
regions, or  more extended  regions possibly  tracing stellar  or disc
winds.  Further analysis  on the precise location  of the Br\,$\gamma$
emission will be  required in order to study  any possible connections
between these two emission lines.

\section{Conclusions}
\label{sec:conclusions}

This   paper  presents  medium   resolution  VLT/X-Shooter   and  high
resolution  VLT/CRIRES near-infrared spectra  of several  Herbig Ae/Be
stars, in an investigation of  the inner regions of their circumstellar
discs.  Below we summarise the main findings:

\begin{itemize}

\item From a  large spectroscopic survey of over 90  HAeBe targets, we
  detect  only  six  objects  exhibiting CO  first  overtone  bandhead
  emission, corresponding to  a detection rate of  approximately 7 per
  cent.  Analysis  of the upper  limits suggests that the  majority of
  non-detections  are not  due  to the  sensitivity  of the  X-Shooter
  instrument.

\item  The  objects displaying  CO  overtone  emission are  mainly  of
  spectral type  B, and are  thus hotter  and more massive  than their
  A-type counterparts.

\item In all  objects that display CO bandhead emission,  we also find
  Br\,$\gamma$  emission of  varying  strengths.  We  find a  positive
  correlation between  the strength of  the CO $v=$ 2--0  bandhead and
  the  Br\,$\gamma$ line,  in agreement  with previous  investigations
  \citep{carr_1989, connelley_2010}, showing  this correlation extends
  to YSOs of higher masses.

\item The  high resolution  spectra of 5  objects exhibiting  CO first
  overtone emission are fitted with  a model of a thin disc undergoing
  Keplerian rotation, and  good fits are obtained to  all spectra.  It
  was  determined  that  the  spectral  resolution  of  the  X-Shooter
  instrument was insufficient to obtain reliable model fits using this
  procedure.

\item The linewidths  of the Br\,$\gamma$ emission  are between 10--40
  times  larger  than the  intrinsic  linewidths  of the  CO  overtone
  emission,  suggesting that  they  originate  in different  kinematic
  environments.

\item The location  of the CO overtone emission in  these best fitting
  models is consistent  with the hypothesis that it  originates from a
  small scale gaseous  disc, interior to the  dust sublimation radius,
  but beyond the co-rotation radius of the central star.

\end{itemize}

It is important  to note that for the object  where spatially resolved
observations   have  also   been  performed,   HD  259431   (MWC  147,
\citealt{kraus_2008a}), we  obtain a  remarkably similar value  to the
inclination  of  the  disc  based  on  our  fitting  technique  ($\sim
50$\degr).  While  this comparison can  currently only be made  in one
object, it does suggest that high spectral resolution observations can
be  used   as  an  alternative  to   interferometric  observations  to
investigate the sub-au scale regions around young stars.

\smallskip

We  plan   to  investigate   this  with  further   observations  using
VLTI/AMBER,  which  will enable  direct  measurements  of the  spatial
extent of the CO-emitting gas,  and allow comparison with our spectral
fitting technique.  This, alongside  more sophisticated modelling that
can include the  vertical structure of inner discs,  will provide much
information on  the nature  of the inner  regions around  Herbig Ae/Be
stars.

\section*{Acknowledgments}

The authors would like to thank  the referee Wing-Fai Thi for comments
that improved  the clarity  of the manuscript.   In addition,  we also
thank Peter Woitke, Rens Waters and  the members of the FP7 DIANA team
for useful discussions.  JDI  gratefully acknowledges funding from the
European  Union  FP7-2011  under  grant agreement  no.   284405.   JRF
gratefully acknowledges a studentship  from the Science and Technology
Facilities Council of the UK.  SK acknowledges support through an STFC
Ernest Rutherford fellowship.

\bibliographystyle{mn2e_long} 

\bsp

\label{lastpage}

\end{document}